\documentclass[preprint,prd]{revtex4}
\newcommand{\bi}{\begin{itemize}}
\newcommand{\ei}{\end{itemize}}

\newcommand{\be}{\begin{equation}}
\newcommand{\ee}{\end{equation}}
\newcommand{\bdm}{\begin{displaymath}}
\newcommand{\edm}{\end{displaymath}}
\newcommand{\bea}{\begin{eqnarray}}
\newcommand{\eea}{\end{eqnarray}}
\newcommand{\nonu}{\nonumber}
\usepackage{epsfig}
\begin{document}
\title{Few exact results on gauge symmetry factorizability on intervals}
\author{Ngoc-Khanh Tran} 
\email[]{nt6b@galileo.phys.virginia.edu}
\affiliation{Department of Physics, University of Virginia \\
382 McCormick Road, Charlottesville, Virginia 22904-4714, USA}
\date{\today}
\begin{abstract}
We track the gauge symmetry factorizability by boundary conditions on  
intervals of any dimensions. With Dirichlet-Neumann boundary 
conditions, the Kaluza-Klein decomposition in five-dimension for 
arbitrary gauge group can always be factorized into that for separate 
subsets of at most two gauge symmetries, and so is completely solvable. 
Accordingly, we formulate a limit theorem on gauge symmetry factorizability
by boundary conditions to recapitulate this remarkable feature of the 
five-dimension case. In higher-dimensional space-time, an interesting 
chained-mixing of gauge symmetries by Dirichlet-Neumann boundary conditions 
is explicitly constructed. The systematic decomposition picture obtained  
in this work constitutes the initial step towards determining the general  
symmetry breaking scheme by boundary conditions.
\end{abstract}
\maketitle
\section{Introduction}
One of the most beautiful piece that lies in the core of  
particle interaction standard model (SM) and elsewhere in condense 
matter physics is the concept of spontaneous symmetry breaking (SSB). In 
particle physics, this so-called Higgs mechanism is capable of generating  
masses for gauge bosons (i.e. breaking the corresponding gauge 
symmetries) and fermions in such a way that renormalizability and 
unitarity can be preserved. However, through extensive experimental search to 
date, the Higgs bosons which are the mechanism's most direct  
proponent have not been observed. In an interesting alternative approach 
with the presence of compact extra spatial dimensions \cite{A}, 
it is known that some gauge bosons of choice may or may not have 
massless mode in 
Kaluza-Klein (KK) decomposition, depending on their assigned boundary 
conditions (BC) and/or transformation properties under certain  
complementary discrete symmetries \cite{Kawamura}. To be specific for 
later referencing, let us list here all three possible Dirichlet-Neumann 
BC configurations for a gauge field $A$ living on an one dimensional 
line segment $0\leq y \leq \pi R$. All of these BCs can be constructed by 
some simple orbifold projection.
\bea
\nonu
&&
\mbox{D-D configuration:}\;\;\;  
A(y)|_{0} =  A(y)|_{\pi R} = 0 \Rightarrow 
A(y)= \sum_{m\in N} A^{(m)}\sin \frac{my}{R} \\
\label{DN}
&&
\mbox{D-N configuration:}\;\;\;  
A(y)|_{0} =  \partial_y A(y)|_{\pi R} = 0 \Rightarrow 
A(y)= \sum_{m\in N} A^{(m)} \sin \frac{(2m+1)y}{2R} \\
&&
\nonu
\mbox{N-N configuration:}\;\;\;  
\partial_y A(y)|_{0} =  
\partial_y A(y)|_{\pi R} = 0 \Rightarrow 
A(y)= \sum_{m\in N} A^{(m)}\cos \frac{my}{R}
\eea
Evidently, only in the N-N configuration, gauge field possesses a 
non-trivial massless ($m=0$) mode. In other words, gauge symmetry is 
preserved still only for boundary conditions at both end-points being 
of Neumann type. This so-called orbifold projection symmetry breaking 
(see e.g. \cite{Kawamura}) has 
distinctive  nature originated truly from boundary effects \cite{HMN}, 
that cannot be identified with the Higgs mechanism through the 
spontaneous breaking of 
background gauge field Wilson line. But again in this orbifold 
compactification approach, up to the possible absence of the zero mode, 
mass spectra of all gauge boson towers are identical, which makes it 
difficult to implement the $SU(2)_L\times U(1)_Y \rightarrow U(1)_{QED}$ 
electroweak breaking/mixing without re-introducing the Higgs below  
compactification scale.

To circumvent this obstacle, more general boundary conditions have been 
proposed for linear combinations of gauge fields rather than for only  
single ones \cite{CGMPT} (earlier models using extra dimensional BCs 
to break the gauge symmetries, with or without traditionally full set 
of Higgs bosons, are described in \cite{AMQ,AB}). As such, the 
corresponding compactification procedure generally grows out of the 
framework restricted by orbifold projections
\footnote{Or else the equivalent set of required projectors, 
if such can be found, 
is quite complicated which makes the projection itself less practical in use.}
, i.e. one is actually working with interval compactification. 
Since BCs are 
imposed on gauge fields' linear combinations, it is plausible that symmetry 
breaking/preserving indeed can be obtained for mixtures of initial 
gauge groups. Then immediately come important queries on the 
systematic/exact connection, if such ever exists, between pattern of 
symmetry breaking/mixing and the BCs being employed. This kind of 
information is 
relevant, because after all, we definitely want to identify those BCs 
that can produce the desired symmetry mixings given beforehand. In the 
case of the Higgs SSB mechanism, it is the Goldstone theorem that can tell all 
about symmetry breaking pattern once the scalar field's vacuum expectation 
value (VEV) configuration is picked. With that in mind, in this work 
we attempt to explore a similar connection between the gauge symmetry 
mixing on one-dimensional flat intervals and the boundary conditions 
applied on their 
ends. The investigation could be eventually generalized for larger number 
of extra dimensions and higher symmetry group ranks. 

To have a taste of what specifically are going to be addressed later on, 
we first briefly review the original Higgsless symmetry breaking 
$SU(2)_L\times SU(2)_R \times U(1)_{B-L} \rightarrow U(1)_{QED}$ on 
a fifth-dimensional interval $0\leq y \leq \pi R$ \cite{CGMPT,CGPT,CGHST} 
(see also \cite{ADMS}). As far as the breaking/mixing is concerned, already, 
this set-up is rich and equally mysterious, so it can serve as the debut 
of the current discussion. After the following Dirichlet and Neumann BCs 
\footnote{These BCs are obtained in \cite{CGMPT} for the limit of a 
brane-localized scalar infinite VEV. There are also BCs on the fifth 
components of gauge fields, which act to eliminate these components 
altogether in an appropriate gauge (section \ref{Propagation}).}
on some combinations of gauge bosons $A_L^{1,2,3}$ (of $SU(2)_L$, 
coupling $g$), $A_R^{1,2,3}$ (of $SU(2)_R$, coupling $g$) and $B$ 
(of $U(1)_{B-L}$, coupling $g'$) are imposed for neutral sector 
\be
\label{NeutralBC} 
\begin{array}{ll}
{\mbox{at $y=0$} 
\left\{ \begin{array}{l}
\left\{ \begin{array}{l}
\partial_y B_{\mu}=0 \\
\partial_y (A_{L\mu}^3 + A_{R\mu}^3) = 0 
\end{array} \right. \\
A_{L\mu}^3 - A_{R\mu}^3 =0
\end{array} \right.} \;\;\;\; ; \;\;\; &
{\mbox{at $y=\pi R$} 
\left\{ \begin{array}{l}
\left\{ \begin{array}{l}
\partial_y A^3_{L_\mu}=0 \\
\partial_y (gB_{\mu} + g'A_{R\mu}^3) = 0 
\end{array} \right.\\
g'B_{\mu} - gA_{R\mu}^3 =0
\end{array} \right.}
\end{array}
\ee
and for charged sector
\be
\label{ChargedBC}
\begin{array}{ll}
{\mbox{at $y=0$} 
\left\{ \begin{array}{l}
\left\{ \begin{array}{l}
\partial_y (A_{L\mu}^1 + A_{R\mu}^1) = 0 \\
\partial_y (A_{L\mu}^2 + A_{R\mu}^2) = 0 
\end{array} \right.\\
\\
\left\{ \begin{array}{l}
A_{L\mu}^1 - A_{R\mu}^1 =0 \\
A_{L\mu}^2 - A_{R\mu}^2 =0
\end{array} \right.
\end{array} \right.}  \;\;\;\; ; \;\;\; &
{\mbox{at $y=\pi R$} 
\left\{ \begin{array}{l}
\left\{ \begin{array}{l}
\partial_y A^1_{L\mu}=0 \\
\partial_y A^2_{L\mu}=0 
\end{array} \right.\\
\\
\left\{ \begin{array}{l}
A^1_{R\mu}=0 \\
A^2_{R\mu}=0 
\end{array} \right.
\end{array} \right.}
\end{array}
\ee
the solutions are found to be (up to normalization factors)
\be
\label{NeutralSol}
\left\{ \begin{array}{l}
B_{\mu}(x,y) = g\gamma_{\mu}(x) + 
g' \sum_{m=1}^{\infty}
\cos {\left( \frac{\pm \phi}{\pi R} + \frac{m}{R} \right)} Z_{\mu}^{(m)}(x)
\\
A^{+3}_{\mu}(x,y)\equiv \frac{A^{3}_{L\mu} + A^{3}_{R\mu} }{\sqrt{2}} = 
g'\gamma_{\mu}(x) - g\sum_{m=1}^{\infty}
\cos{\left( \frac{\pm \phi}{\pi R} + \frac{m}{R} \right)} Z_{\mu}^{(m)}(x) 
\\
A^{-3}_{\mu}(x,y)\equiv \frac{A^{3}_{L\mu} - A^{3}_{R\mu} }{\sqrt{2}} = 
g\sum_{m=1}^{\infty}
\sin{\left( \frac{\pm \phi}{\pi R} + \frac{m}{R} \right)} Z_{\mu}^{(m)}(x) 
\\
\;\;\;\;\;\;\;\;\;\; \mbox{(where $\tan \phi \equiv \sqrt{g^2+2g'^2}/g$)}
\end{array} \right.
\ee
\be
\label{ChargedSol}
\left\{ \begin{array}{l}
A^{+1,2}_{\mu}(x,y)\equiv \frac{A^{1,2}_{L\mu} + A^{1,2}_{R\mu} }{\sqrt{2}} = 
\sum_{m=1}^{\infty}
\cos{\left( \frac{\pm 1}{4 R} + \frac{m}{R} \right)} W_{\mu}^{1,2(m)}(x) 
\\
A^{-1,2}_{\mu}(x,y)\equiv \frac{A^{1,2}_{L\mu} - A^{1,2}_{R\mu} }{\sqrt{2}} = 
\sum_{m=1}^{\infty}
\sin{\left( \frac{\pm 1}{4 R} + \frac{m}{R} \right)} W_{\mu}^{1,2(m)}(x)
\end{array} \right.
\ee
Beyond the obvious fact that the solutions 
(\ref{NeutralSol}), (\ref{ChargedSol}) 
satisfy the BCs (\ref{NeutralBC}), (\ref{ChargedBC}), it is tempting to gain 
more insights from this construction. 

How comes only the intended $U(1)_{QED}$ 
(of massless photon $\gamma_{\mu}(x)$) is kept unbroken globally along the 
entire interval, following the different breakings 
$SU(2)_L\times SU(2)_R \times U(1)_{B-L} \rightarrow SU(2)_D \times U(1)_{B-L}
$ and  $SU(2)_L\times SU(2)_R \times U(1)_{B-L} \rightarrow 
SU(2)_L \times U(1)_{Y}$ made at $y=0$ and $y=\pi R$ respectively. This 
is because, as indicated in (\ref{DN}), the combination of (\ref{NeutralBC}) 
allows uniquely a $U(1)$ gauge field to have Neumann BC at both end-points. 
A further question then is whether it is possible to track the symmetries 
that might be left unbroken/broken locally at any given point 
$y\neq 0, \pi R$ in the bulk, for the eventual 
sake of model building with brane-localized matter fields. Next, looking at 
the neutral sector alone, one begins with three gauge degrees of freedom 
($A_L^3(x,y),A_R^3(x,y),B(x,y)$) in 5-dim, nevertheless it appears that 
one ends up with only one 
Kaluza-Klein (KK) tower ($\{Z^{(k)}(x)\}$) 
and eventually two different mass spectra in 4-dim, and a similar observation 
can be made for charged sector. This is seemingly because, as expected from 
differential 
equation theory, the BCs (\ref{NeutralBC}) (or (\ref{ChargedBC})) render 
the ansatz that gives the mixed solutions and may constrain the number of free 
parameters in them. A further question then is how effectively BCs can be used 
to limit the number of 4-dim field towers, for the eventual sake of symmetry  
group rank reduction as in some classes of grand unified theory (GUT) 
breaking. Next, 
we also see from (\ref{NeutralSol}), (\ref{ChargedSol})
that the numbers of Neumann (and Dirichlet) BCs are equal on two ends of 
the interval, but these quantities do not represent the number of unbroken 
(and broken) gauges as naively expected form Eqs. (\ref{DN}). 
This is because, as 
mentioned above, the BCs induce a mixing between initial group's gauge fields, 
and what might matter for the KK decomposability is perhaps the equality 
of BCs' total number ($N+D$) on end-points. A further question then is how  
the mixing/breaking forms if, say, there are $D$ Dirichlet, $N$ Neumann BCs 
at $y=0$, and $D'$, $N'$ ones at $y=\pi R$ with a single constraint 
$D+N=D'+N'$, for the eventual sake of generalizing the construction. 

In this paper we seek to answer these and discuss some other general  
questions on flat interval compactification with arbitrary gauge 
group and viable boundary conditions beyond Dirichlet 
and Neumann types. The recipe being exclusively used here to track 
the broken symmetry  is to find out the gauge sector KK spectra under 
the given  BCs, and identify their possible massless modes.      
In section \ref{Propagation}, we reconstruct in details the solutions 
(\ref{NeutralSol}), (\ref{ChargedSol}) using geometrical arguments. 
Essentially, each solution of a definite topological (KK) number
can be regarded as a map between two boundary 
sets of constant fields. In this view, we cannot only track how the symmetry 
breaking propagates in the bulk, but also show that the expressions 
(\ref{NeutralSol}), (\ref{ChargedSol}) are not yet the 
most general solutions obeying the BCs (\ref{NeutralBC}), (\ref{ChargedBC}). 
Rather, photon and $Z$-boson can have different KK towers 
(this observation has been also made in \cite{CGPT}). In section 
\ref{Factorizability}, 
we consider any large set of gauge fields living in a fifth dimensional 
interval, which obey any Dirichlet-Neumann BC configurations. It is shown 
that the general gauge space can always be factorized, by virtue of 
Dirichlet-Neumann BC, into mutually orthogonal subspaces of no more than two 
dimensions each. Surprisingly, in this way, the initial general set-up 
is broken into smaller ones, which are solvable and indeed not more 
complicated than the original Higgsless set-up described above. 
In section \ref{From} we analyze the orthogonality 
between massive gauge field modes as well as the compatibility between 
D-N BCs and variational principle of action. This is important for 
the construction of 4-dim effective Lagrangian and the phenomenologies 
that follow it, such as pertubative unitarity in gauge bosons 
scattering \cite{LQT}. Although {\em a priori} the fore-mentioned 
orthogonality is 
not apparent for system with mixed symmetry breaking under consideration, 
{\em a posteriori} both orthogonality and normalization of KK towers that 
survive in 4-dim are particularly transparent and simple.  
In section \ref{HigherDim} we give an explicit and systematic construction
of gauge symmetry decomposition in higher dimension, which reveals an 
interesting and strict ``chained entanglement'' from one symmetry to 
the other by BCs. The possibility to non-trivially mix up to $2^d$ 
gauge symmetries in $d$ extra dimensions is also demonstrated.
In section \ref{Conclusion} we summarize the main results with some 
outlooks, and finally 
in the Appendix A we prove a lemma on matrix factorizability, which is  
the basis for gauge space factorization presented in section 
\ref{Factorizability}.

Throughout the presentation, we repeatedly invoke simple geometric 
interpretation/visualization to support our arguments. This approach  
is particularly suited for the flat space considered here. The method also 
appears very helpful for warped space investigation. 
\section{Gauge symmetry breaking along an 1-dim interval}
In this section, we will consider a pure gauge set-up in $4+1$ space-time. 
The fifth dimension is finite, and the respective 
coordinate $y$ runs from $0$ to $\pi R$. 
\subsection{Propagation of symmetry breaking}
\label{Propagation}
Before tackling the original Higgsless model described in the introduction 
section, let us work out first an apparently simpler problem with just 
two gauge degrees of freedom $N(x,y)$, $D(x,y)$. It turns out 
that this is all we practically need for the solution of the gauge 
multi-dimensional problem, as long as the Dirichlet-Neumann BCs are being 
employed. The action of this set-up is 
\be
\label{action}
{\cal S}=\sum_{A=N,D}\int d^4 x\int_0^{\pi R} dy 
\left( -\frac{1}{4} F^{A}_{\mu\nu}F^{A\mu\nu}
-\frac{1}{2} (\partial_{\mu} A_{5} - \partial_{5} A_{\mu})
(\partial^{\mu} A^{5} -\partial^{5} A^{\mu})
+ \ldots + {\cal L}_{GF} \right)
\ee
where $F^{A}_{\mu\nu} \equiv \partial_{\mu}A_{\nu} - \partial_{\mu}A_{\nu}$ 
is the linearized field tensors, while the dots represent possible triple and 
quartic interactions for non-Abelian groups. As in gauge theory in 5-dim  
space-time, the 4-dim covariant gauge fixing term 
${\cal L}_{GF}=\frac{1}{2\xi}
(\partial_{\mu} A_{\mu} - \xi \partial_{5} A_{5})^2$ is chosen to render the 
cancellation of bilinear mixing between $A_{\mu}$ and $A_5$. However, for 
the interval compactification, such cancellation is only up to a non-trivial 
surface term
\be
\label{surface}
\left.\int_0^{\pi R} dy \left( -\frac{1}{2} F^{A}_{5\mu}F^{A5\mu} + 
\frac{1}{2\xi}(\partial_{\mu} A_{\mu} - \xi \partial_{5} A_{5})
(\partial^{\mu} A^{\mu} - \xi \partial^{5} A^{5}) \right) \supset
(\partial_{\mu} A^{\mu})A^5 \right|_0^{\pi R}
\ee
which is not necessarily always zero for any $A_5$ configuration. 
Nevertheless, in unitary gauge $\xi \rightarrow \infty$ the derivative 
$\partial_{5} A_{5}$ in ${\cal L}_{GF}$ tends to zero, i.e. $A_5$ may have 
only massless mode. We further can choose either Dirichlet-Dirichlet or 
Dirichlet-Neumann BC configurations (\ref{DN}) for $A_5$ to suppress this  
zero mode. Thus in the case of 5-dim space-time, we can make all gauge field 
fifth components disappear by choice of gauge and BCs, and also 
eliminate the residual surface term (\ref{surface}). A more careful 
consideration 
\cite{MPR} shows that in the unitary gauge, all non-zero KK modes 
$A_5^{(m)}$ indeed are ``eaten'' by longitudinal modes of massive bosons 
$A_{\mu}^{(m)}$. Thus hereafter, we can take $A_5=0$ identically.

We impose the following BCs on different set of gauge fields $\{N,D\}$ at 
$y=0$ and $\{N',D'\}$ at $y=\pi R$ (with $0\leq \phi <  \pi$)
\be
\label{2BC}
\left\{
\begin{array}{l}
\partial_y N_{\mu}(x,y)|_{y=0} = 
\partial_y N'_{\mu}(x,y)|_{y=\pi R}=0 \\
D_{\mu}(x,y)|_{y=0} = D'_{\mu}(x,y)|_{y=\pi R}=0
\end{array} \right.
\;\;\;
\mbox{where}
\;\;\;
\left( \begin{array}{l}
N' \\
D'
\end{array} \right) \equiv
\left( \begin{array}{cc}
\cos \phi & -\sin \phi \\
\sin \phi &  \cos \phi
\end{array} \right)
\left( \begin{array}{l}
N \\
D
\end{array} \right)
\ee
Obeying the linearized motion equations derived from 
the action (\ref{action}), all gauge fields can be expressed in term of 
combination of trigonometric functions. Further, the BCs (\ref{2BC}) 
allow us to look for ${N,D}$ in the form: $N(x,y)= N(x)\cos{My}$, 
$D(x,y)=D(x)\sin{My}$. Then from the rewritten BCs on ${N',D'}$ at $y=\pi R$ 
\be
\begin{array}{c}
\partial_y N'_{\mu}(x,y)|_{y=\pi R} = -M
\left[ N_{\mu}(x)\cos\phi\sin{(M\pi R)} + D_{\mu}(x)\sin\phi\cos{(M\pi R)} 
\right] = 0 \\
D'_{\mu}(x,y)|_{y=\pi R} = N_{\mu}(x)\sin{\phi} \cos{(M\pi R)} +
D_{\mu}(x)\cos{\phi} \sin{(M\pi R)}=0
\end{array}
\ee
one finds
\be
\label{1spectrum}
N(x)=\pm D(x) \;\;\;\;\mbox{and}\;\;\;\; \tan{M\pi R} =\mp \tan{\phi} 
\Rightarrow M=\frac{m}{R} \mp \frac{\phi}{\pi R} \;\; (m \in Z)
\ee
There seem to be two mass spectra corresponding to two different signs in 
Eq. (\ref{1spectrum}). However, as the integer $m$ runs from minus to plus 
infinity, the two spectra can be merged into one 
(i.e. $|M^{(m)}_{-}|=|M^{(-m)}_{+}|$), which allows to write the general 
solution of fields in the following 
form
\be
\label{1sol}
\begin{array}{l}
N(x,y)= \sum_{m \in Z}  N^{(m)}(x) 
\cos{\left(\frac{m}{R} + \frac{\phi}{\pi R}\right)y}  \\
D(x,y)= \sum_{m \in Z}  - N^{(m)}(x) 
\sin{\left(\frac{m}{R} + \frac{\phi}{\pi R}\right)y}  \\
N'(x,y)= \sum_{m \in Z}  N^{(m)}(x) 
\cos{\left((\frac{m}{R} + \frac{\phi}{\pi R})y - \phi\right)} \\ 
D'(x,y)= \sum_{m \in Z} - N^{(m)}(x) 
\sin{\left((\frac{m}{R} + \frac{\phi}{\pi R})y - \phi\right)}  
\end{array}
\ee
Indeed the BCs generate a relation (\ref{1spectrum}) between the coefficients  
$N(x)$, $D(x)$, and thus produce only a ``single'' independent KK tower 
of gauge boson in 4-dim. However, for general $\phi$, this tower 
(with $m \in Z$) 
has twice as many 4-dim modes as does the tower (with $m \in N$) given in 
Eq. (\ref{DN}) for $\phi = 0$ or $\frac{\pi}{2}$
\footnote{In Eq. (\ref{DN}), because extra dimensional wavefunctions 
are identical (up to a sign) for a pair of negative and positive  
$m$, we can combine them and the resulting sum is limited to $m\in N$. 
In contrast, because of the presence of a general twist angle $\phi$, 
such consolidation is not possible for (\ref{1sol}), thus $m\in Z$ therein.}
. So properly speaking, the 
single tower $\{N^{(m)}(x)\}$ in Eq. (\ref{1sol}) will be referred to as an 
extended tower hereafter. We choose to adopt this notation to simplify the 
writing and to eliminate the possible ambiguity over the double sign $\pm$ in 
(\ref{1spectrum}). We note that the lowest mass depends on $\phi$: 
it is $\frac{\phi}{\pi R}$ (or $m=0$) for $0 \leq \phi \leq \frac{\pi}{2}$, 
and 
$\frac{\pi-\phi}{\pi R}$ (or $m=-1$) for $\pi > \phi > \frac{\pi}{2}$. 
Finally, 
when $\phi=0$ or $\frac{\pi}{2}$, we have $(N\sim N';D\sim D')$ or 
$(N\sim D';D\sim N')$ respectively, i.e. $N$ decouples from $D$ and each 
forms an independent 1-dim system with known BCs at both end-points 
given in (\ref{DN}). The gauge symmetry associated with $N$ is 
unbroken.

\begin{figure}
        \begin{center}    
        \epsfig{figure=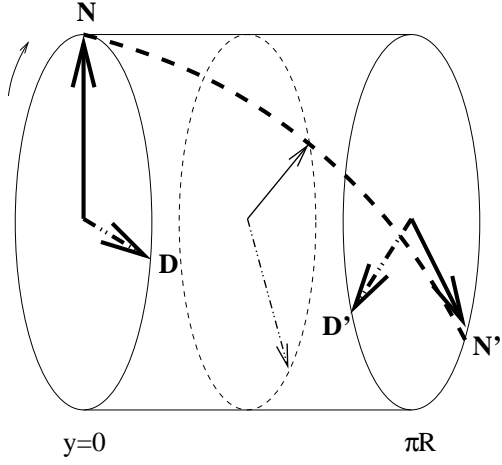,width=0.4\textwidth}
        \end{center}
        \vspace{0cm}
        \caption{Clockwise mapping $\{N,D\}$ at $y=0$ to $\{N',D'\}$ at 
$y=\pi R$. The thick dashed line shows the bulk trajectory of Neumann 
direction in gauge symmetry space of some specific mode with map's 
winding number $m$.}
        \label{Cl}
        \end{figure}

\begin{figure}
        \begin{center}    
        \epsfig{figure=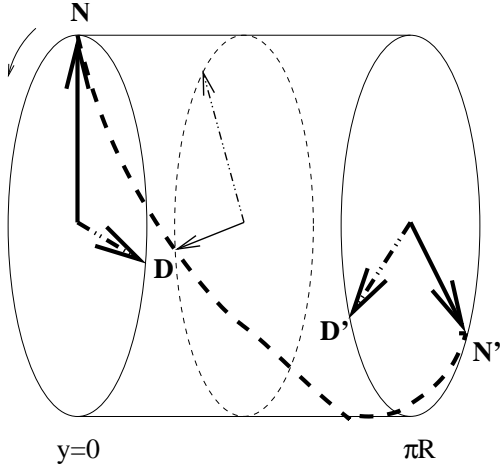,width=0.4\textwidth}
        \end{center}
        \vspace{0cm}
        \caption{Counter-clockwise mapping $\{N,D\}$ at $y=0$ to $\{N',D'\}$ 
at $y=\pi R$. The thick dashed line shows the bulk trajectory of Neumann 
direction in gauge symmetry space of some specific mode with map's 
winding number $m$.}
        \label{AtCl}
        \end{figure}

This solution may be seen better geometrically. First, the definition 
(\ref{2BC}) of the set $\{N',D'\}$ implies that it is rotated from the 
orthogonal set $\{N,D\}$ 
\footnote{The quadratic terms of Lagrangian (\ref{action}) are orthogonal in 
$\{N,D\}$ basis, so these can be seen as two orthogonal eigenvectors 
in gauge space.} 
by an angle $\phi$, while the latter satisfying Dirichlet-Neumann  
BCs can be conveniently cast in the form 
\be
\label{map}
\left(\begin{array}{l}
N(x,y) \\ D(x,y)
\end{array} \right)  = 
\left(\begin{array}{cc}
\cos{My} & -\sin{My} \\
\sin{My} & \cos{My}
\end{array} \right)
\left(\begin{array}{c}
N(x) \\ 0
\end{array} \right)
\ee  
Because this describes a $O(2)$-rotation, it is also suggestive to pretend  
that, in term of value, the set $\{N(x,y),D(x,y)\}$ in turn is rotated from 
its original $\{N(x),0\}$ (at $y=0$) by the angle $My$ as it propagates 
in the bulk. Next, the BCs (\ref{2BC}) essentially signal that, again in 
term of values, $\{N',D'\}$ at $y=\pi R$, up to a sign, should be identical 
to $\{N,D\}$ at $y=0$, which can be met when two above rotations precisely 
cancel one another. In expression, this is (see also Figs. (\ref{Cl}), 
(\ref{AtCl})), 
\be
\begin{array}{rcl}
\label{2spectra}
\mbox{for clockwise rotation:} \;\;\;
& M\pi R= \phi+m\pi \;\;\; 
&(m\in Z) \\
\mbox{for counter-clockwise rotation:} \;\;\;
& M\pi R= -\phi+m\pi \;\;\;
&(m\in Z) 
\end{array}
\ee
Clearly, KK mode number $m$ is none other than the number of 
(half-)revolutions that the map from $\{N(x,y),D(x,y)\}$ at $y=0$ to 
$\{N'(x,y),D'(x,y)\}$ at $y=\pi R$ wraps around, whether clockwise or 
counter-clockwise. Each value of $m$ presents 
a solution for which the given BCs hold. But we again note that 
a rotation of $m$ (half-)revolutions clockwise is equivalent to that of 
$-m$ ones counter-clockwise, so indeed the two spectra in (\ref{2spectra}) 
are the same. Hence the complete solution (\ref{1sol})  
consists of only a single (extended) KK towers $\{N^{(m)}(x)\}$ of  
4-dim gauge fields. This geometrical 
construction further might shed light into the pattern of symmetry 
breaking/mixing at any point in the bulk. If we could limit the 
consideration to a particular solution ($m$ fixed), it then follows  
from Eq. (\ref{map}) that, in gauge space, the locally preserved 
symmetry direction (with vanishing field derivative) at point $y$ would 
make an angle $My$ with respect to vector field $N$, while the completely 
broken symmetry direction (with vanishing field) would make an angle 
$My+\frac{\pi}{2}$. 
Accordingly, if zero mode ($M=0$) exists, then in that mode the locally 
preserved symmetry direction is unchanged ($My=0$) for all $y$, i.e. 
that well-defined symmetry can be said to be unbroken throughout the bulk. 
This observation is in line with the effective 4-dim picture  
(where the fifth coordinate is integrated out leaving no mass term for 
zero mode), 
because zero mode wave function is constant on the entire interval. In 
either descriptions, the KK zero mode always is a reliable indicator of the 
associated preserved symmetry. 
However, because these rotation angles change with general $m$, when  
the comprehensive solution (summed over all $m$) (\ref{1sol}) is undertaken 
in an intact 5-dim view, no where in the bulk a fixed direction can 
represent either locally preserved or completely broken symmetry, with the 
exception of two end-points. 
 
\begin{figure}
        \begin{center}    
        \epsfig{figure=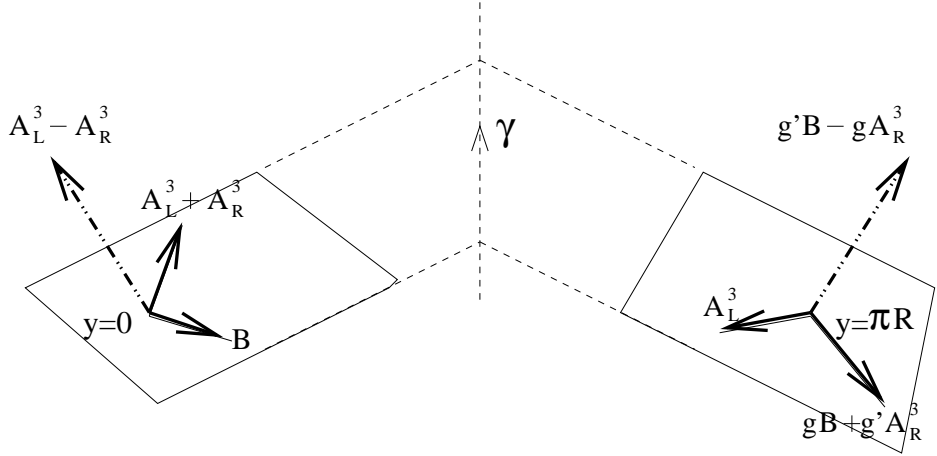,width=0.75\textwidth}
        \end{center}
        \vspace{0cm}
        \caption{Two 2-dim Neumann planes (defined at $y=0$ and 
$y=\pi R$ respectively) intersect in gauge symmetry space along a 1-dim 
line, which is the $U(1)_{QED}$ unbroken symmetry direction with 
massless photon $\gamma$.}
        \label{21-1}
        \end{figure}

\begin{figure}
        \begin{center}    
        \epsfig{figure=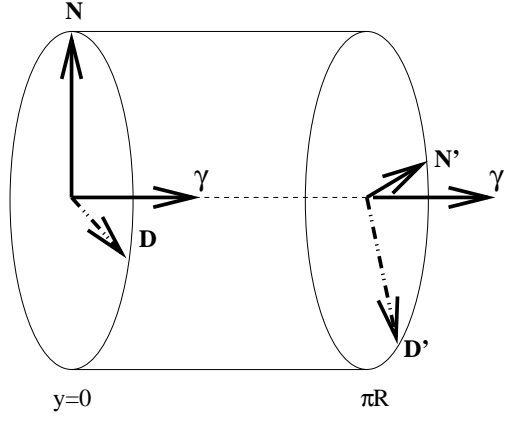,width=0.4\textwidth}
        \end{center}
        \vspace{0cm}
        \caption{Relative orientation of gauge fields at $y=0$ and $y=\pi R$.}
        \label{21-2}
        \end{figure}

We are now ready to embark on the original Higgsless configuration. The 
neutral sector $\{A_{3L},A_{3R},B\}$ presents a 3-dim gauge space, with 1-2 
(Dirichlet-Neumann) BCs on each ends. Since both Neumann and Dirichlet BCs 
are closed under additivity, at either end-points, we can have 
non-intersecting Dirichlet and 
Neumann subspaces, each contains only gauge fields satisfying the respective 
BC. In the Higgsless 3-dim neutral gauge space 
(\ref{NeutralBC}), the 2-dim  
Neumann subspace $\{B,\frac{A_{L}^3 + A_{R}^3}{\sqrt{2}}\}$ at $y=0$ and 
the other 2-dim Neumann subspace 
$\{A_L^3, \frac{gB + g'A_{R}^3}{\sqrt{g^2+g'^2}}\}$ at 
$y=\pi R$ generally share one co-dimension, just exactly as two 2-dim planes 
intersect along a 1-dim line in 3-dim space (Fig. \ref{21-1}). 
That is, there exists one gauge 
vector field satisfying the Neumann BC at both end-points, i.e. by virtue of 
(\ref{DN}), the associated $U(1)_{QED}$ gauge symmetry is unbroken throughout 
the interval. It is straightforward to obtain this preserved symmetry 
direction $\gamma$, and then the orthogonal-to-it remaining 1-1 
(Dirichlet-Neumann) spaces at end-points (Fig. \ref{21-2}). Note also that we 
generally cannot further factorize these residual 2-dim spaces, because 
the co-dimension of their constituent Dirichlet (or Neumann) subspaces, 
one at $y=0$ and the other at $y=\pi R$, is zero, just exactly as 
two 1-dim lines intersect at a dimensionless point in 2-dim plane.   
\be
\label{2sol}
\begin{array}{c}
\mbox{at $y=0$ }
\left\{
\begin{array}{ll}
\partial_y \gamma_{\mu}|_{y=0} = 0; &
\gamma\equiv \frac{g'(A_{3L}+A_{3R})+gB}{\sqrt{g^2+2g'^2}} \\
\partial_y N_{\mu}|_{y=0} = 0; &
N\equiv \frac{g(A_{3L}+A_{3R})-2g'B}{\sqrt{2g^2+4g'^2}}  \\
D_{\mu}|_{y=0} = 0; &
D\equiv \frac{A_{3R}-A_{3L}}{\sqrt{2}} 
\end{array} \right.
\\
\mbox{at $y=\pi R$ }
\left\{
\begin{array}{ll}
\partial_y \gamma_{\mu}|_{y=\pi R} = 0; &
\gamma\equiv \frac{g'(A_{3L}+A_{3R})+gB}{\sqrt{g^2+2g'^2}}  \\
\partial_y N'_{\mu}|_{y=\pi R} = 0; &
N'\equiv 
\frac{(g^2+g'^2)A_{3L} -g'^2 A_{3R}-gg'B}{\sqrt{(g^2+g'^2)(g^2+2g'^2)}} \\
D'_{\mu}|_{y=\pi R} = 0; &
D'\equiv \frac{gA_{3R}-g'B}{\sqrt{g^2+g'^2}} 
\end{array}  \right.
\end{array} 
\ee
Both $\{\gamma,N,D\}$ and $\{\gamma,N',D'\}$ sets are explicitly 
orthonormal, however $\{N',D'\}$ is rotated from $\{N,D\}$ by an angle $\phi$ 
(see Fig. \ref{21-2} and Eq. (\ref{2BC})).
\be
\label{2ND}
\left( \begin{array}{l}
N' \\
D'
\end{array} \right) 
= 
\left( \begin{array}{cc}
\frac{g}{\sqrt{2g^2+2g'^2}} 
& -\frac{\sqrt{g^2+2g'^2}}{\sqrt{2g^2+2g'^2}} \\
\frac{\sqrt{g^2+2g'^2}}{\sqrt{2g^2+2g'^2}}
& \frac{g}{\sqrt{2g^2+2g'^2}}
\end{array} \right)
\left( \begin{array}{l}
N \\
D
\end{array} \right)
\Rightarrow \tan{\phi}=\frac{\sqrt{g^2+2g'^2}}{g}
\ee
The solutions of the ``decoupled'' $\gamma(x,y)$ and the ``entangled'' 
pair $\{N,D\}$ now are readily given in (\ref{DN}) and (\ref{1sol}) 
\be
\label{3sol}
\begin{array}{l}
\gamma(x,y)= \gamma^{(0)}(x) + 
\sum_{m\neq 0}\cos{\frac{my}{R}} \gamma^{(m)}(x) \\
N(x,y)= \sum_{m \in Z}  Z^{(m)}(x) 
\cos{\left(\frac{m}{R} + \frac{\phi}{\pi R}\right)y}  \\
D(x,y)= \sum_{m \in Z} -Z^{(m)}(x) 
\sin{\left(\frac{m}{R} + \frac{\phi}{\pi R}\right)y} 
\end{array}
\ee
where $\phi$ is given in Eq. (\ref{2ND}) and we have changed the notation 
from $N(x)$ in (\ref{1sol}) to $Z(x)$ in (\ref{3sol}) to facilitate the 
comparison with the earlier Higgsless solution. 
From (\ref{3sol}) one can obtain $\{B,A_{3L},A_{3R}\}$ (again up to 
normalization factors) after using the definitions in (\ref{2sol})
\bea
\nonu
B_{\mu}(x,y)  = 
g \gamma^{(0)}_{\mu}(x) + 
g\sum_{m=1}^{\infty} \gamma^{(m)}_{\mu}(x) \cos{\frac{my}{R}} 
-g' \sum_{m\in Z}
Z_{\mu}^{(m)}(x)\cos {\left( \frac{m}{R} + \frac{ \phi}{\pi R} \right)y}
\eea
\bea
\label{NeutralSol1}
A^{+3}_{\mu}(x,y) && \equiv \frac{A^{3}_{L\mu} + A^{3}_{R\mu} }{\sqrt{2}}  = 
g'\gamma^{(0)}_{\mu}(x) + 
\sum_{m=1}^{\infty} g' \cos{\frac{my}{R}} \gamma^{(m)}_{\mu}(x)  \\
\nonu
&&
+ g\sum_{m \in Z}  Z_{\mu}^{(m)}(x) 
\cos{\left( \frac{m}{R} + \frac{ \phi}{\pi R} \right)y} 
\eea
\bea
\nonu
A^{-3}_{\mu}(x,y) \equiv \frac{A^{3}_{L\mu} - A^{3}_{R\mu} }{\sqrt{2}} = 
g\sum_{m \in Z}  Z_{\mu}^{(m)}(x) 
\sin{\left( \frac{m}{R} + \frac{ \phi}{\pi R} \right)y} 
\eea
where $\phi$ is given in (\ref{2ND}). For the charged sector 
$\{A^{1}_{L},A^{2}_{L},A^{1}_{R},A^{2}_{R}\}$, we 
first split it into two 
separate subspaces generated respectively by $\{A^{(1)}_{L},A^{(1)}_{R}\}$ 
and $\{A^{(2)}_{L},A^{(2)}_{R}\}$. Each of these 2D subspaces is 
just the familiar case of  
1-1 Dirichlet-Neumann BCs (\ref{2BC}) with $\phi=\frac{\pi}{4}$, so their 
solutions follow immediately from (\ref{1sol})
\be
\label{ChargedSol1}
\begin{array}{l}
A^{+1,2}_{\mu}(x,y)\equiv \frac{A^{1,2}_{L\mu} + A^{1,2}_{R\mu} }{\sqrt{2}} = 
\sum_{m\in Z} 
W_{\mu}^{1,2(m)}(x)
\cos{\left(  \frac{m}{R}+ \frac{ 1}{4 R} \right)y}  
\\
A^{-1,2}_{\mu}(x,y)\equiv \frac{A^{1,2}_{L\mu} - A^{1,2}_{R\mu} }{\sqrt{2}} = 
\sum_{m\in Z} 
W_{\mu}^{1,2(m)}(x)
\sin{\left( \frac{m}{R} + \frac{1}{4 R}  \right)y}  
\end{array} 
\ee
The expressions (\ref{NeutralSol1}), 
(\ref{ChargedSol1}) are more general than (\ref{NeutralSol}), 
(\ref{ChargedSol}) obtained first in \cite{CGMPT}
\footnote{This fact has also been noted in \cite{CGPT}}
, because here photon $\gamma^{(0)}(x)$ generally belongs 
to a distinctive (factorized) KK tower $\{\gamma^{(m)}(x)\}$ of 
distinctive mass spectra $\{\frac{m}{R}\}$. The mass 
$\frac{\phi}{\pi R}$ of $Z$-boson zero mode is always lighter than 
$\frac{1}{R}$ of photon's first excited state. Thus the presence of photon 
KK tower does not spoil the ability to mimic the SM spectrum of this original  
Higgsless model found in \cite{CGMPT} at low energy scale. It needs, however, 
be taken into account in the processes involving real or virtual 
massive gauge bosons, for e.g., to check the model's unitarity. Further, we 
present (\ref{NeutralSol1}), (\ref{ChargedSol1}) in the
single-spectrum form at the price of letting $m$ have both negative and 
positive integral values.

We have just seen that, indeed the BCs can reduce half of the number 
of independent 4-dim gauge field towers 
\footnote{Again note that the tower left in 4-dim picture is an 
``extended'' tower.} 
in a simple set of two ``entangled'' gauge symmetries  
$\{N,D\}$. We next will explore this ability of Dirichlet-Neumann BCs, 
which are imposed again on the fifth interval's two ends, but for  
arbitrary number of gauge symmetries.
\subsection{Factorizability for arbitrary gauge dimensions}
\label{Factorizability}
In this section, let us assume that, at $y=0$, there be $D$ and $N$ gauge 
fields obeying the Dirichlet and Neumann BCs  respectively. At $y=\pi R$, 
let those quantities be $D'$ and $N'$. These gauge fields may represent 
Abelian (like $B(x,y)$) or non-Abelian (like $A^{3}_{L,R}$) 
symmetries. As we are interested in obtaining the KK decomposition of 
these fields, we just work first with the linearized motion equations
\footnote{It is important to note that, the non-linear terms may 
recombine after the decomposition to produce complex symmetry breaking 
patterns. The investigation of this non-abelian recombination is crucial in
determining the general symmetry breaking schemes by boundary conditions. 
This study, however, lies beyond the scope of the current work.}. 
Therefore, within the stated purpose, 
there is no difference between 
Abelian and non-Abelian treatment. We also suppose that each of $\{N+D\}$ 
and $\{N'+D'\}$ be an orthonormal set of gauge vectors, just as 
naturally as we 
are working with gauge eigenstates that appear in the initial 5-dim 
Lagrangian. 
The original Higgsless set-up obviously belongs to this class 
of construction. Finally, above BCs are perceived to act on four usual 
gauge field components, since in the $4+1$ space-time, the fifth components 
can be made identically vanished by appropriate gauge's choice and 
BCs on them.

In the case $D+N \neq D'+N'$, after eventual basis transformations, 
there will be fields with BCs being specified whether at only one of 
two end-points or more than once at a same end-point. The former situation 
generally leads to a non-quantized spectrum because of insufficient BCs, while 
the latter could lead to no spectrum at all because of redundant 
(and conflicting) BCs. In this work we will not pursue either of these 
directions, though some particular BCs might be carefully selected 
to evade these shortcomings in the construction of standard KK decomposition.

In the remaining case $D+N = D'+N'$, we take, without loss of generality, 
$D>N,N'$. We can always assume further that each of the 
following pairs $(\{N\},\{N'\})$, $(\{N\},\{D'\})$, $(\{N'\},\{D\})$ are 
non-intersecting, i.e.  
$codim(\{N\},\{N'\})=codim(\{N\},\{D'\})=codim(\{N'\},\{D\})=0$ 
\footnote{Notation: $codim(\{N\},\{D'\})$ denotes the co-dimension of Neumann 
subspace generated by $\{N\}$ gauge fields and the Dirichlet subspace 
generated by $\{D'\}$ gauge fields in gauge space.}, 
because if these are not zero, we can find (and solve) 
a number of gauge fields, each has one of three BCs configurations listed in 
Eq. (\ref{DN}). After we decouple these ``solvable'' fields 
from the set-up (just as we decouple $U(1)_{QED}$ gauge field from the 
Higgsless neutral sector in previous section), we are left with only 
non-intersecting gauge boson sets. However, these sets are not necessarily 
mutually orthogonal in general (just as $\{N,D\}$ and $\{N',D'\}$ in Eq. 
(\ref{2BC}) generally make an angle $\phi \neq  \frac{\pi}{2}$).

We first create the union space $(\{N\} \cup \{N'\})$, which  
has $N+N'$ dimensions, since $codim(\{N\},\{N'\})=0$. Next, we construct 
the space $\{D-N'\}$ of dimension $D-N'$ that complements  
$(\{N\} \cup \{N'\})$ in the entire gauge space, i.e. 
\bea
\nonu
\{D-N'\} \cup (\{N\} \cup \{N'\}) = \{D+N\} 
\eea
Moreover, this 
complementary space $\{D-N'\}$ can always be built in such a way that 
it is mutually orthogonal to both $\{N\}$ and $\{N'\}$, that is any 
vector in $\{D-N'\}$ is orthogonal to all vectors in $\{N\}$ 
and $\{N'\}$, and vice versa. 

As such, $\{D-N'\}$ necessarily is a subspace of $\{D\}$, because 
by construction, $\{D\}$ is the largest space that is mutually orthogonal 
to $\{N\}$ in the entire gauge space being spanned by the orthonormal 
set $\{N+D\}$. Similarly, $\{D-N'\}$ also necessarily forms a subspace of 
$\{D'\}$, which implies
\bea
\nonu
\{D-N'\} \subset (\{D\} \cap \{D'\})
\eea  
In other words, each of $D-N'$ gauge fields that generate the space 
$\{D-N'\}$ has Dirichlet BC at both  $y=0$ (because it belongs to $\{D\}$) 
and $y=\pi R$ (because it belongs to $\{D'\}$). The KK 
decomposition of these $D-N'$ fields is given in (\ref{DN}), according to 
which all $D-N'$ associated symmetries are broken, and on the way we obtain 
$D-N'$ independent gauge boson massive KK towers in 4-dim of identical mass 
spectrum $\{\frac{m}{R}\}$ (with integer $m>0$). 

But this is not the end of the ``decoupling'' process. We are currently  
left with $N+N'$ gauge symmetries, among them 
$N'$ and $N$ orthonormal fields satisfy respectively the Dirichlet and 
Neumann BCs at $y=0$. At $y=\pi R$ those numbers are $N$ (for Dirichlet's) 
and $N'$ (for Neumann's). For clarity, let us denote the corresponding spaces 
generated by these gauge vector sets as $\{N'_{D,0}\}$, 
$\{N_{N,0}\}$, $\{N_{D,\pi R}\}$ and $\{N'_{N,\pi R}\}$, where the subscripts  
are informative about the type and position of BCs, while 
the main capital letters are about the number of fields as always. 
Again, without loss of generality we take $N'>N$ and repeat the above 
steps to further identify and isolate the ``decoupled'' sectors. 
We construct the space $\{N'-N\}$ of $N'-N$ dimensions that complements 
the union space $(\{N_{N,0}\}\cup \{N_{D,\pi R}\})$ of $2N$ dimensions 
(because $codim(\{N_{N,0}\},\{N_{D,\pi R}\})=0$) in 
the (now) entire gauge space $\{N+N'\}$, i.e.
\bea
\nonu
\{N'-N\} \cup (\{N_{N,0}\}\cup \{N_{D,\pi R}\}) = \{N+N'\} 
\eea
Since $\{N'-N\}$ are chosen to be mutually orthogonal to both 
$\{N_{N,0}\}$ and $\{N_{D,\pi R}\}$, it must be a subspace of the 
intersection $(\{N'_{D,0}\} \cap \{N'_{N,\pi R}\})$. These $N'-N$ fields 
then have Dirichlet BC at $y=0$ and Neumann's at $y=\pi R$. In the result 
all the $N'-N$ associated gauge symmetries are necessarily broken. Their KK 
decomposition (\ref{DN}) consists of $N'-N$ independent massive gauge boson  
towers in 4-dim of identical mass spectrum $\frac{2m+1}{2R}$.

At this point, after two successive (but similar) reduction processes, 
there are still $2N$ ``coupled'' gauge symmetries by the following BCs 
distribution (in the above notation): $\{N_{D,0}\}$, $\{N_{N,0}\}$, 
$\{N_{D,\pi R}\}$ and $\{N_{N,\pi R}\}$. Each of   
$(\{N_{D,0}\} \cup \{N_{N,0}\})$ and $(\{N_{D,\pi R}\} \cup \{N_{N,\pi R}\})$ 
forms an orthonormal basis of the (now) entire $2N$-dim gauge space and so 
one is necessarily related to the other 
by some general orthogonal ``BC-defining'' matrix $O(2N)$ 
(i.e. the matrix given in the BCs definition like Eq. (\ref{2BC})). In 
practice, knowing fields on which BCs are imposed (e.g. 
Eq. (\ref{NeutralBC})) we can determine the BC-defining matrix (e.g. 
Eq. (\ref{2ND})). Repeating the same simple argument 
once more likely would not 
help to advance the symmetry ``disentanglement'', if there ever is such 
possibility for the $2N$-dim space at hands. There exists, however, a 
convincing clue that indeed this gauge space is subject to further 
``disentanglement''. Because fields obeying Dirichlet (or Neumann) 
BCs form a closed set under their linear combination transformations, along 
the decoupling process, we have the freedom to perform 
four independent $O(N)$ rotations separately on the sets $\{N_{D,0}\}$, 
$\{N_{N,0}\}$, $\{N_{D,\pi R}\}$ and $\{N_{N,\pi R}\}$. This implies that, 
among the $\frac{2N(2N-1)}{2}$ parameters of the initial BC-defining 
matrix $O(2N)$, 
we can eliminate $4\frac{N(N-1)}{2}$ ones by four $O(N)$ basis 
transformations. We are left with $N$ physical parameters (say 
$\{\phi_1,\ldots,\phi_N\}$) which exactly equals 
the number of parameters in a set of $N$ $O(2)$ matrices. We then 
vaguely expect that the general $2N$-symmetry set could decouple at least
into $N$ independent elementary sets, each consists of no more than two gauge 
dimensions $\{N,D\}$ discussed in details in section IIA. Interestingly, 
this expectation turns out to be a rigorous result obtained through a 
lemma introduced in the Appendix \ref{Alemma}, which confirms the 
diagonal factorizability 
of any general $O(2N)$ matrix into $N$ $O(2)$ blocks by four $O(N)$ 
independent rotations. In this manner, for a general (non-zero) set 
$\{\phi_1,\ldots,\phi_N\}$, the initial $2N$ 
symmetries are all broken by BC compactification, in the result of 
which there are $N$ distinctive gauge field KK towers in 4-dim of 
distinctive mass spectra $\{\frac{m}{R}+\frac{\phi_i}{\pi R}\}$ 
(with $i=1$ to $N$). Again, this factorization can be 
visualized geometrically as follows. Given at the onset two arbitrary 
orthonormal eigenbases ($\{N_{D,0} \cup N_{N,0}\}$ and 
$\{N_{D,\pi R} \cup N_{N,\pi R}\}$) 
in $2N$-dim spaces, one can rearrange internally the basis vectors within 
four $N$-dim subsets ($\{N_{D,0}\}, \{N_{N,0}\}, \{N_{D,\pi R}\}, 
\{N_{N,\pi R}\}$) to divide the initial space into $N$ mutually 
orthogonal 2-dim planes. Each of these planes can be 
identified by two alternative bases made of the fore-mentioned rearranged 
vectors. For $i$-th plane, the two bases are off one from the other by 
angle $\phi_{i}$, which characterizes the corresponding KK mass spectrum. 
In the special case $\phi_{i}=0$ for some $i \in [0,N]$, the 
corresponding $i$-th plane can be further decoupled into 2 independent 
directions, one of which generally represents a preserved symmetry 
as shown in section \ref{Propagation}.

From the proof presented in the Appendix, we in particular note two 
important points. First, the factorization is unique once the 
``BC-defining'' matrix $O(2N)$ is given, i.e. the KK decomposition spectra  
are unambiguously determined for the set of the initial $2N$ Dirichlet-Neumann 
BCs. Second, the fact that $N$ $2$-dim planes obtained in the factorization 
process are {\em mutually orthogonal} is an indispensable condition for them 
to develop $N$ associated {\em independent} spectra. Or putting in another 
order, this statement means: any subspace (with known BCs) being mutually 
orthogonal to the rest can be factored out, and becomes a separate 
problem subject to KK decomposition on its own right.

Let us now recapitulate our discussion so far on the pattern of symmetry 
breaking by BCs and the resulting spectra in the following {\em limit theorem 
on gauge symmetry factorizability by D-N BCs}:
 
{\em For a set of $S$ gauge 
symmetries on a fifth dimensional interval subject to totally $2S$ Dirichlet 
and Neumann BCs, (i) the total number of independent (extended) gauge 
field towers (and equally of associated  mass spectra) in 4-dim is no 
less than $\frac{S}{2}$, (ii) the number of unbroken symmetries
(and equally of mass spectra with zero mode) is generally given by the 
number of mutually orthogonal gauge fields that 
have Neumann BCs at both end-points}.

\begin{figure}
        \begin{center}    
        \epsfig{figure=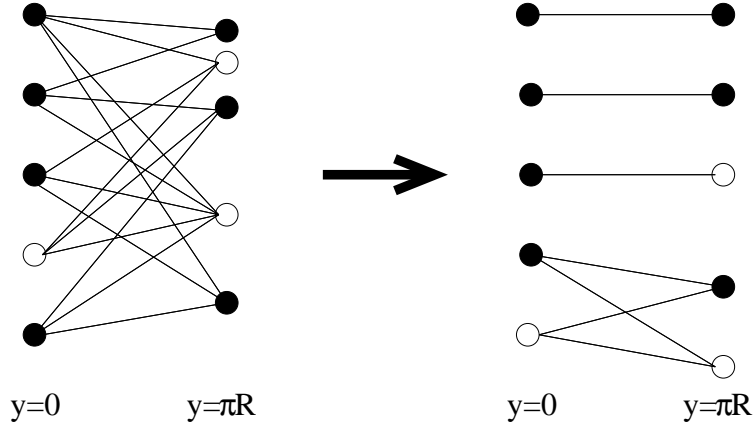,width=0.6\textwidth}
        \end{center}
        \vspace{0cm}
        \caption{Gauge symmetry sets before (left scheme) and after 
(right scheme) some basis transformations (or factorization). 
Shaded and empty dots represent gauge fields with Neumann and Dirichlet BC 
respectively. Connected fields are related (``entangled'') in their 
definition. Within each scheme, fields being in the same horizontal level 
are identical.}
        \label{factoriazation}
        \end{figure}

The more quantitative description of this {\em theorem} is presented in the 
table I for the situation where the non-abelian recombination 
(see footnote $[$30$]$) 
does not hold, like in the simple original Higgsless set-up 
\cite{CGMPT} studied above. 
\begin{table}
\caption{Symmetry breaking on interval with Dirichlet-Neumann boundary 
conditions.}
\begin{tabular}{|c|c|c|c|}
\hline
Total initial symmetries & \multicolumn{3}{c|}{$S=D+N=D'+N'$} \\
\hline
\begin{tabular}{c}
BC distribution \\
(\# Dirichlet, \# Neumann)
\end{tabular}
& 
\multicolumn{3}{c|}{$(D,N)_{ y=0 }\; \& \; (D',N')_{y=\pi R}$} \\
\hline
\begin{tabular}{c} 
Assumption \\
(without generality loss)
\end{tabular}
& \multicolumn{3}{c|}{ 
\begin{tabular}{c} 
$D>N'>N$ \\
$codim(\{N\},\{N'\})=codim(\{N\},\{D'\})=codim(\{N'\},\{D\})=0$ 
\end{tabular} }
\\ \hline
Sector's symmetry & Completely broken & Completely broken & 
Mixed broken \& unbroken \\
\hline 
Sector's dimensions & $D-N'$ & $N'-N$ & $2N$ \\
\hline
\begin{tabular}{c}
BC type \\
($y=0$)-($y=\pi R$)
\end{tabular}
& Dirichlet-Dirichlet & Dirichlet-Neumann & Mixed \\
\hline
Mass spectrum & $\frac{m}{R}\;\; (m\neq 0)$ & $\frac{2m+1}{2R}$ 
& $\frac{m}{R}+ \frac{\phi_i}{\pi R}$ \\
\hline
\# 4-dim independent towers & $D-N'$ & $N'-N$ & $\geq N$ \\
\hline
\end{tabular}
\end{table}
The specific case $N=1,D=4,N'=2,D'=3$ is illustrated in Fig. 
\ref{factoriazation}.
While the above {\em theorem} literally identifies the number of 4-dim towers 
into which the fields' modes are grouped as the result of 
compactification, it is  
really about the {\em limit} to ``entangle'' the symmetries by 
Dirichlet-Neumann BCs. In case of a single extra dimension, one can at 
most combine two initial gauge directions to produce a single 
(though extended) tower of vector fields, which we see in 4-dim. 
Evidently, this ``hybrid'' tower receives contribution from both initial 
5-dim gauge fields, and this observation is crucial for the modes' 
orthogonality and normalization as we discuss next. When more extra 
dimensions are allowed, we will see in section \ref{HigherDim} that more 
fields can be ``entangled''.
\section{From KK mode orthogonalities to effective Lagrangian in 4-dim}
\label{From}
In this section we will construct the 4-dim effective Lagrangian that results 
from integrating out the fifth dimension. The integration gives rise to 
surface terms, that a priori do not automatically vanish because fields 
with defined BCs are not the same on the two boundaries. Rather, the 
nullifying of these terms sets constraints on ``qualified'' BCs through 
the general variational principle on the action. In section 
\ref{orthogonality} we examine the KK mode orthogonality from both onset 
BCs and KK decomposition. In section \ref{normalization} 
we perform the explicit normalization 
of extra dimensional wavefunctions and thus obtain the effective gauge boson 
self-couplings.    
\subsection{Orthogonality between KK modes}
\label{orthogonality}
We first consider the (real) KK wavefunctions $g^{(k)}(y),g^{(l)}(y)$ 
corresponding to modes $k,l$ of an initial gauge field $G(x,y)$ in 5-dim  
($\partial_y^2 g^{(k)}= m_k^2 g^{(k)}$; 
$\partial_y^2 g^{(l)}= m_l^2 g^{(l)}$). From 
the integration by parts
\be
\int_{0}^{\pi R} g^{(k)} \partial_y^2 g^{(l)} dy 
= g ^{(k)}\partial_y g^{(l)} |_{0}^{\pi R} 
- (\partial_y g^{(k)}) g^{(l)} |_{0}^{\pi R}
+ \int_{0}^{\pi R} (\partial_y^2 g^{(k)})g^{(l)} dy
\ee 
follows the wavefunction overlap
\be
\label{overlap}
\left.
(m_l^2 - m_k^2) \int_{0}^{\pi R} g^{(k)} g^{(l)}dy = 
[g^{(k)} \partial_y g^{(l)} -(\partial_y g^{(k)}) g^{(l)} ]
\right|_{0}^{\pi R}
\ee 
For $k=l$ this equality trivially holds, but for $k \neq l$,   
if we impose the orthogonality between $\{g^{(k)}\}$, this is translated 
into a condition on the BCs
\be
\label{orthocon}
\left.
[g^{(k)} (\partial_y g^{(l)}) -(\partial_y g^{(k)}) g^{(l)} ]
\right|_{0}^{\pi R} = 0
\ee 
In practice, one may have BCs on either $G$ or $\partial_y G$ (but not both)  
at an end-point, or else their BCs are not known at the other end-point 
(like the Dirichlet-Neumann set-up of previous section),  
so a priori the orthogonality condition (\ref{orthocon}) is not evidently 
met by itself in interval compactification. Actually, the subtle 
cancellation holds only collectively between all fields in each ``entangled''  
sector as we will see now. It is by the way worthwhile to note that 
(\ref{orthocon}) is equivalent to the completeness relation given 
in \cite{CGMPT} (because both have same root in 
the hermiticity of $\hat{\partial}_y^2$ in a finite interval), which is 
required to annihilate the $E^4$ term in longitudinal gauge boson 
scattering amplitude (see also \cite{CDH,CDHN}). We will come back to this 
assertion in the next section.

From the previous analysis, any gauge symmetry set with Dirichlet-Neumann 
BCs can be factorized into decoupled subsets of one gauge dimension and 
``entangled'' subsets of two dimensions. The decoupled subsets 
(e.g. $\gamma(x,y)$ of neutral sector in Eq. (\ref{2sol})) 
have as BCs one of three configurations listed in Eq. (\ref{DN}), so the 
orthogonality for decoupled sector is obviously satisfied. As a direct 
check for the KK decomposition (\ref{DN}), we indeed have 
\be
\int_0^{\pi R} \sin{\frac{ky}{R}}\sin{\frac{ly}{R}} \, dy \sim
\int_0^{\pi R} \sin{\frac{(2k+1)y}{2R}}\sin{\frac{(2l+1)y}{R}} \, dy \sim
\int_0^{\pi R} \cos{\frac{ky}{R}}\cos{\frac{ly}{R}} dy \sim 
\delta_{k,l}
\ee
because $k,l \in N$. If one repeats this direct check for KK 
decomposition of field $D(x,y)$ (or $N(x,y)$) give in Eq. (\ref{1sol})
\be
\label{nonortho}
\begin{array}{l}
\int_0^{\pi R} \sin{\left(\frac{k}{R} + \frac{\phi}{\pi R} \right)y}
\sin{\left(\frac{l}{R} + \frac{\phi}{\pi R}\right)y} \;dy \neq 0  \\
\int_0^{\pi R} \cos{\left(\frac{k}{R} + \frac{\phi}{\pi R}\right)y}
\cos{\left(\frac{l}{R} + \frac{\phi}{\pi R}\right)y}\; dy \neq 0
\end{array}
\ee
one finds apparently KK modes of $D(x,y)$ (or $N(x,y)$) are 
not orthogonal to one the others. This can also be explained by the fact 
that orthogonality 
condition (\ref{orthocon}) does not hold for each of $D(x,y)$ and $N(x,y)$ 
as suspected. This seems to be a little counter-intuitive to the standard 
KK decomposition procedure. Either in continuous \cite{MPR} or  
latticized \cite{HS,H-G,FGS,CHKST1,CHKST2,H,CHKST3} (see also \cite{CH}) 
extra dimension models with presence 
of some brane/bulk scalar VEVs, the orthogonality of gauge bosons modes 
is quasi-automatic because these are eigenmodes of a hermitian matrix to be 
diagonalized. In the current situation, as $D(x,y)$ and $N(x,y)$ 
belong to an entangled sector and specially share the same 4-dim tower 
$\{N^{(m)}(x)\}$, we should not have looked at each of them individually. 
Rather, let us check the KK mode orthogonality and simultaneously derive  
the effective Lagrangian of the sector $\{D(x,y),N(x,y)\}$ as a whole. 
We again concentrate on the abelianized part of the action and set all 
field fifth components to zero, leaving non-quadratic interactions 
as well as normalization factors for the next section's effective couplings 
discussion.   
\bea
\nonu
{\cal L}^{(4)}= && \sum_{k,l\in Z} \int_0^{\pi R} dy 
\left( -\frac{1}{4} F^{N(k)}_{\mu\nu}F^{N(l)\mu\nu}
-\frac{1}{4} F^{D(k)}_{\mu\nu}F^{D(l)\mu\nu}
-\frac{1}{2}  F^{N(k)}_{5\mu}F^{N(l)5\mu}
-\frac{1}{2}  F^{D(k)}_{5\mu}F^{D(l)5\mu}
\right)  \\ \label{ortho}
= && \sum_{k,l\in Z} 
\left(-\frac{1}{4} F^{(k)}_{\mu\nu}(x) F^{(l)\mu\nu} (x) 
+\frac{1}{2} (\frac{k\pi+\phi}{\pi R}) (\frac{l\pi+\phi}{\pi R}) 
N^{(k)}_{\mu}(x) N^{(l)\mu}(x) \right)  \\ \nonu
\times &&
\int_0^{\pi R} dy
\left( \cos{\frac{(k\pi+\phi)y}{\pi R} } \cos{\frac{(l\pi+\phi)y}{\pi R} } 
+ \sin{\frac{(k\pi+\phi)y}{\pi R}} \sin{\frac{(l\pi+\phi)y}{\pi R} } 
\right) 
\eea
where the 4-dim linearized field tensor 
$F^{(k)}_{\mu\nu}(x) \equiv \partial_{\mu} N^{(k)}_{\nu}(x)- 
\partial_{\nu} N^{(k)}_{\mu}(x)$ and we have used the explicit solution 
(\ref{1sol}). Evidently, the integration over the fifth 
coordinate produces a Kronecker's delta $\delta_{k,l}$, thus enforces the 
orthogonality between modes $N^{(k)}_{\mu}(x)$ in 4-dim. From Eqs. 
(\ref{nonortho}), (\ref{ortho}) it is clear that this 
orthogonality can be achieved only when all 5-dim fields of an entangled 
sector are taken into consideration, because indeed the sector's 
unique 4-dim tower is part of all of them. Such visualization might also serve 
in favor of the 4-dim ``extended'' tower viewpoint adopted for each 
independent sector.

Alternatively, we now can see more directly the role of BC on 
the KK modes' orthogonality from another perspective. This time we work 
with the Dirichlet-Neumann BCs, but not the expilicit decomposition solution. 
When the extra dimension is integrated out to obtain
the effective Lagrangian, we note 
that only the cross-terms containing the derivative over 
fifth coordinate ($\partial_y N_{\mu} \in F_{5\mu}$) can generate  
surface contribution to the 4-dim action
\bea
\label{45}
&&{\cal S}^{(4)} \supset -\frac{1}{2}  
\int_0^{\pi R} dy
\left( F^{N}_{5\mu}(x,y) F^{N5\mu}(x,y) + 
F^{D}_{5\mu}(x,y)F^{D5\mu}(x,y) \right) =
\\ \nonu
&&\frac{1}{2} \left. \left[ \{N_{\mu}(x,y) \; D_{\mu}(x,y)\} 
\left\{ \begin{array}{c}
\partial_y N_{\mu}(x,y)  \\
\partial_y D_{\mu}(x,y)
\end{array} \right\}  
\right|_{0}^{\pi R}
- \int_0^{\pi R} dy \{ N_{\mu}(x,y) \; D_{\mu}(x,y)\} 
\left\{ \begin{array}{c}
\partial_y^2 N_{\mu}(x,y) \\
\partial_y^2 D_{\mu}(x,y)
\end{array} \right\}
\right]
\eea 
Because all gauge fields obeying the Laplace's equation in the bulk, 
we can replace 
$\partial_y^2$ in the second term of (\ref{45}) by some 4-dim squared mass 
$M^{2}$, and then this term's orthogonality follows in exactly the same way  
as it does for the pure 4-dim kinetic terms considered earlier. 
In contrast, the surface term could present some unwanted contribution 
(which might spoil the mode orthogonality), and in model-building practice 
should be set to zero by appropriate choice of BCs. Indeed, with 
Dirichlet-Neumann BC, the surface term of effective action (\ref{45}) 
vanishes identically at both end points and thus it leaves no destructive
effect on the orthogonality of KK decompositon in 4-dim. 
 
\subsection{Normalization and effective gauge boson self-couplings}
\label{normalization} 
Let us now complete the construction of effective Lagrangian by 
normalizing the extra dimensional wave functions. As we have already seen, 
the mode diagonality concerns all symmetries altogether within each 
independent sector, and normalization should be proceeded in the coherent 
way. That is, one should not normalize $N$ and $D$ modes separately basing 
on the expansion (\ref{1sol}).

For the decoupled sector (one of three configurations in Eq. (\ref{DN})), 
the normalization factor is $\sqrt{2/\pi R}$ for all modes, with 
the only exception (being $\sqrt{1/\pi R}$) of the massless mode in the 
Neumann-Neumann configuration. For the entangled sector (\ref{2BC}), the 
normalization is given by the integration given in the sector's  
effective Lagrangian (\ref{ortho}) (putting $l=k$)
\be
\left[
\int_0^{\pi R} dy
\left( \cos{\frac{(k\pi+\phi)y}{\pi R} } \cos{\frac{(l\pi+\phi)y}{\pi R} } 
+ \sin{\frac{(k\pi+\phi)y}{\pi R}} \sin{\frac{(l\pi+\phi)y}{\pi R} } 
\right) \right]^{-1/2}_{(k=l)}= \frac{1}{\sqrt{\pi R}} 
\ee
It is interesting to note that the normalization factor is independent of 
both mode number and twist angle $\phi$. 
As an illustration, we are now ready to write down the complete 
(normalized) KK expansions 
(\ref{NeutralSol1}), (\ref{ChargedSol1}) of the original Higgsless model
\be
\label{completesol}
\begin{array}{c}
B(x,y)  = \frac{g \gamma^{(0)}(x) + g\sqrt{2}
\sum_{m\neq 0} 
\gamma^{(m)}(x)\cos{\frac{my}{R}}  
-g'\sqrt{2}\sum_{m\in Z}
Z^{(m)}(x)\cos {( \frac{m}{R} + \frac{ \phi}{\pi R})y}}
{\sqrt{\pi R}\sqrt{g^2+2g'^2}} \\
A^{3}_{L,R}(x,y) = \frac{g' \gamma^{(0)}(x) + g'\sqrt{2}\sum_{m\neq 0} 
\gamma^{(m)}(x)\cos{\frac{my}{R}}  
+ \sqrt{g^2+g'^2}\sum_{m\in Z}
Z^{(m)}(x)\cos \{( \frac{m}{R} + \frac{ \phi}{\pi R})y \mp \phi \}}
{\sqrt{\pi R}\sqrt{g^2+2g'^2}} \\
A^{1,2}_{L}(x,y) = \frac{\sum_{m\in Z}
W^{1,2(m)}(x) \cos \{( \frac{m}{R} + \frac{ 1}{4 R})y - \frac{\pi}{4} \}}
{\sqrt{\pi R}}  \\ 
A^{1,2}_{R}(x,y) = \frac{\sum_{m\in Z}
W^{1,2(m)}(x) \sin \{( \frac{m}{R} + \frac{ 1}{4 R})y - \frac{\pi}{4} \}}
{\sqrt{\pi R}} 
\end{array}
\ee
and the respective effective Lagrangian (with the $SU(2)$ complete 
field tensor 
$F^{a}_{MN}(x,y) = \partial_M A^{a}_N(x,y) - \partial_N A^{a}_M(x,y) + 
g\epsilon^{abc}A^{b}_M(x,y) A^{c}_N(x,y) $ and 5-space-time Lorentz 
indices $M,N=0,\ldots,4$)
\bea
\nonu
{\cal L}^{(4)}  = && \frac{-1}{4}\int_{0}^{\pi R} 
\left( F^{B}_{MN} F^{BMN} + 
\sum_{a=1}^{3}F^{La}_{MN} F^{LaMN} + \sum_{a=1}^{3}F^{Ra}_{MN} F^{RaMN}
\right) dy  \\ \nonu
= && \left(
\sum_{A=\gamma,Z,W^1,W^2} \sum_m 
\frac{-1}{4}(\partial_{\mu}A^{(m)}_{\nu} - \partial_{\nu}A^{(m)}_{\mu})
(\partial^{\mu}A^{(m)\nu} - \partial^{\nu}A^{(m)\mu}) 
+\frac{1}{2}M_A^{(m)} A^{(m)}_{\mu}A^{(m)\mu} \right) \\ \label{completeL}
&&- \frac{g\sin{\phi}\cos{\phi}\sqrt{g^2+g'^2}}{\sqrt{\pi R(g^2+2g'^2)}}
\left( \frac{1}{\phi}-\frac{1}{2\phi-\pi}-\frac{1}{2\phi+\pi} \right) 
(\partial_{\mu} W^{1(0)}_{\nu})W^{2(0)\mu} Z^{(0)\nu} \\ \nonu
&&- \frac{gg'}{\sqrt{\pi R(g^2+2g'^2)}}
(\partial_{\mu} W^{1(0)}_{\nu})W^{2(0)\mu} \gamma^{(0)\nu}  
+ \ldots
\eea
where the dots denote the triple and quartic gauge interactions, which can 
be found by plugging the expansions (\ref{completesol}) into the effective 
Lagrangian. Whereas, we have explicitly specified the zero mode triple 
interactions $W^{1(0)}W^{2(0)}Z^{(0)}$ and $W^{1(0)}W^{2(0)}\gamma^{(0)}$ 
in Eq. (\ref{completeL}). If $(W^{1(0)}\mp iW^{2(0)})/\sqrt{2}$ are to be 
identified with SM charged bosons $W^{\pm}$, then the ratio of these 
couplings presents the SM weak mixing angle
\be
\frac{g\sin{\phi}\cos{\phi}\sqrt{g^2+g'^2}}{g'}
\left( \frac{1}{\phi}-\frac{1}{2\phi-\pi}-\frac{1}{2\phi+\pi} \right)
= \cot{\theta_W} 
\ee
Since $\tan{\phi}=\frac{\sqrt{g^2+2g'^2}}{g}$, this gives a  
relation between 5-dim couplings $g,g'$. We can also straightforwardly 
compute other effective couplings between gauge field higher modes. In 
particular, the self-interaction of $W^{1(m)} W^{2(n)} \gamma^{(p)}$ is 
found to be
\be
g_{W^{1(m)} W^{2(n)} \gamma^{(p)}}= 
g'\sqrt{2}\int_0^{\pi R} dy \cos{\frac{py}{R}}\cos{\frac{(m-n)y}{R}}
\sim \delta_{p+m-n,0} + \delta_{p-m+n,0}
\ee
This effective coupling signals an ``accidental'' fifth-momentum 
conservation, a 
result of the precise canceling contribution from the initial 
$SU(2)_{L}$ and $SU(2)_{R}$ to $W$ towers. This raises a possibility 
(which is highly model-dependent) that, due to symmetry entanglement by BCs, 
pertubative unitarity could be achieved in some carefully constructed 
models when the criterion (\ref{orthocon}) is not held 
separately for each initial field.

Though matter fields 
(Higgs scalars and fermions) can be introduced at either 
end-points or throughout the bulk, it is not easy to make this original 
Higgsless set-up in 5-dim flat space-time cope simultaneously with other 
SM constraints \cite{CGMPT}. More realistic versions of this model  
take into account the warped geometry factor \cite{CGPT} or more extra 
dimensions \cite{GNS} (see also next section). To encompass these 
directions, a more general 
study of gauge space factorizability in interval compactification is 
underway and will be presented elsewhere \cite{inpre}.
\section{Perspectives on gauge space symmetry mixing in higher dimensions}
\label{HigherDim}
By now, the consideration of section \ref{Factorizability} has led to the 
conclusion that in 5-dim space, one cannot ``entangle'' more than 
two gauge symmetries by however-sophisticated Dirichlet-Neumann BCs. A 
natural question to ask is whether this limitation on gauge mixing by BCs 
can be somewhat lifted in higher-dim space (that is, can we mix more 
symmetries with more dimensions?). The final answer is affirmative, and in 
this section we will demonstrate this general 
trend through an explicit, but systematic and simple construction of 
symmetry mixing in any dimensions. 
\subsection{Few simplest extensions}
Let us first consider the simplest extension of the above 5-dim set-up:  
we now work with 3-dim gauge problem $\{G_1,G_2,G_3\}$ in the 6-dim space 
(i.e. two extra dimensions) with the coordinates denoted 
as $(x_{\mu},y_1,y_2)$.
The extra dimensions are all finite with length $\pi R$ 
(i.e. $0\leq y_1,y_2 \leq \pi R$).  
There exist different ways to define the boundary conditions on this 
extra dimensional square. Here again, the observation of 
section \ref{Factorizability} can help simplify our choice. Along 
1-dim interval, say along $Oy_1$ (i.e. $y_2=0$), one can truly mix only  
two out of three gauge bosons $\{G_1,G_2,G_3\}$. This prompts us to define the 
now-familiar, but indeed most general, Dirichlet-Neumann BCs on 
the following field sets  
along $Oy_1$ direction (see also Eq. (\ref{2BC}))
\bea
\nonu
&&\mbox{ along $Oy_1$:}
\\
&&
\label{Y1}
\left\{
\begin{array}{c}
\partial_{y_1} G_1 |_{y_1=0} = \partial_{y_1} G'_1 |_{y_1=\pi R} =0 \\
G_2 |_{y_1=0} =  G'_2 |_{y_1=\pi R} =0 
\end{array}
\right. \;\;\;
\mbox{where} \;\;\;
\left(
\begin{array}{c}
G'_1 \\ G'_2 \\ G'_3
\end{array}
\right) \equiv
\left(
\begin{array}{ccc}
\cos \phi_1 & -\sin \phi_1 & 0 \\
\sin \phi_1 &\cos \phi_1 & 0 \\
0&0&1
\end{array}\right)
\left(
\begin{array}{c}
G_1 \\ G_2 \\ G_3
\end{array}
\right)
\eea
In the above equation, $\{G'_1,G'_2,G'_3\}$ is just an auxiliary field set  
(an alternative basis of the same 3-dim gauge space) which serves to specify 
the BCs, and the gauge field $G_3$ represents the 
``disentangled'' symmetry along $Oy$, for which we do not need to identify 
the BC. Below, we will explain this point after we obtain the decomposition
solution for the gauge fields. Similarly, another set of BCs can be imposed
along $Oy_2$ in the same logical pattern
\bea
\nonu
&&\mbox{ along $Oy_2$:}
\\
\label{Y2}
&&
\left\{
\begin{array}{c}
\partial_{y_2} G_2 |_{y_2=0} = \partial_{y_2} G''_2 |_{y_2=\pi R} =0 \\
G_3 |_{y_2=0} =  G''_3 |_{y_2=\pi R} =0 
\end{array}
\right. \;\;\;
\mbox{where} \;\;\;
\left(
\begin{array}{c}
G''_1 \\ G''_2 \\ G''_3
\end{array}
\right) \equiv
\left(
\begin{array}{ccc}
1&0&0 \\
0&\cos \phi_2 & -\sin \phi_2  \\
0&\sin \phi_2 &\cos \phi_2  
\end{array}\right)
\left(
\begin{array}{c}
G_1 \\ G_2 \\ G_3
\end{array}
\right)
\eea
The decomposition solution of each of (\ref{Y1}), (\ref{Y2}) separately
has been readily obtained in Eq. (\ref{1sol}). However, to accommodate both 
(\ref{Y1}), (\ref{Y2}) we need to add in certain multiplicative 
factors $h(y_1)$ and $l(y_2)$ 
\be
\label{G1G2}
\left\{
\begin{array}{l}
G_1\sim  \cos(\frac{m_1}{R}+\frac{\phi_1}{\pi R})y_1 \;l(y_2)  \\
G_2\sim -\sin(\frac{m_1}{R}+\frac{\phi_1}{\pi R})y_1 \;l(y_2)
\end{array}
\right.
\ee
\be
\label{G2G3}
\left\{
\begin{array}{l}
G_2\sim  h(y_1)\cos(\frac{m_2}{R}+\frac{\phi_2}{\pi R})y_2 \\
G_3\sim -h(y_1) \sin(\frac{m_2}{R}+\frac{\phi_2}{\pi R})y_2
\end{array}
\right.
\ee
where $m_1,m_2$ are all integers. 
After consolidating $G_2$ in (\ref{G1G2}) and (\ref{G2G3}) we can solve for 
$h(y_1)$, $l(y_2)$ and unambiguously 
obtain the general decomposition of the gauge fields from our 6-dim set-up
\bea
\nonu
G_1(x_{\mu},y_1,y_2)&=&\sum_{m_1,m_2\in Z}g^{(m_1,m_2)}(x_{\mu})\;
\cos{\left([\frac{m_1}{R}+\frac{\phi_1}{\pi R}]y_1\right)}\;
\cos{\left([\frac{m_2}{R}+\frac{\phi_2}{\pi R}]y_2\right)}
\\
\label{6sol}
G_2(x_{\mu},y_1,y_2)&=&-\sum_{m_1,m_2\in Z}g^{(m_1,m_2)}(x_{\mu})\;
\sin{\left([\frac{m_1}{R}+\frac{\phi_1}{\pi R}]y_1\right)}\;
\cos{\left([\frac{m_2}{R}+\frac{\phi_2}{\pi R}]y_2\right)}
\\
G_3(x_{\mu},y_1,y_2)&=&\sum_{m_1,m_2\in Z}g^{(m_1,m_2)}(x_{\mu})\;
\sin{\left([\frac{m_1}{R}+\frac{\phi_1}{\pi R}]y_1\right)}\;
\sin{\left([\frac{m_2}{R}+\frac{\phi_2}{\pi R}]y_2\right)}
\nonu
\eea
There are a number of interesting observations we can make by looking at 
the above decomposition. First, the general solution (\ref{6sol}) 
clearly indicates that the composite KK tower $g^{(m_1,m_2)}(x)$ of gauge 
boson in 4-dim is synthesized from all three 
initial gauge fields $G_1,G_2,G_3$ in 6-dim. That is, in other words, 
in higher dimensions we can indeed mix/break more than two symmetries by 
general boundary conditions. Second, when 
$\phi_y=\phi_z=\phi_u=0$,   
it comes with no surprise that in this model, $G_2, G_3$ 
do not have non-trivial zero mode, and we expect that the associated 
symmetries are completely broken in 4-dim effective picture. This is 
because in Eqs. (\ref{Y1}), (\ref{Y2}), $G_2,G_3$  do have Dirichlet BC at 
certain boundaries. 

Third and most amazingly, in the transition from 
the disconnected decompositions $G_1-G_2$ and $G_2-G_3$
(\ref{G1G2}),(\ref{G2G3}) to the composite solutions $G_1-G_2-G_3$ 
(\ref{6sol}), gauge symmetries $G_1$ and $G_3$ have been ``entangled'', 
though we did not start out with any explicit BCs that mix these two 
symmetries. Mathematically, this ``chained-entanglement''
is manifested by the fact that the 
identification of $G_2$ in (\ref{G1G2}),(\ref{G2G3}) unambiguously 
determines both coefficient $h(y_1)$ and $l(y_2)$. Had we imposed another 
D-N boundary condition of the type (\ref{Y1}) or (\ref{Y2}) on $G_1-G_3$ 
sector, we would have ended up generally with no decomposition solution at 
all by the redundancy of BCs on the underlying (differential) motion 
equation of fields (see the 3-extra dim. set-up below). In a more visual 
description, the first 
BC entangles $G_1$ with $G_2$, the second BC entangle $G_2$ with $G_3$, and 
in the result all three are automatically inter-connected. 

Now we can also see better the reason why in the above construction we did not 
specify the BCs for $G_3$ along 
$y_1$, and $G_1$ along $y_2$, respectively in (\ref{Y1}) and (\ref{Y2}). 
Explicitly, Eq. (\ref{Y2}) states the BCs on 
$G_2$, $G_3$ only along $y_2$, but {\em implicitly}, Eq. (\ref{Y2}) also 
forcefully requires that the analytic dependences of $G_2$ and $G_3$ wave 
functions on coordinate $y_1$ be identical, otherwise BCs (\ref{Y2}) 
simply do not hold. (In Eq. (\ref{G2G3}) above, 
we used the same 
expression $h(y_1)$ in both $G_2(y_1,y_2)$ and $G_3(y_1,y_2)$ just to 
enforce this implicit effect of (\ref{Y2}) boundary conditions). Furthermore, 
Eq. (\ref{Y1}) explicitly imposes BC on $G_2$ along $y_1$. Thus, through 
the virtue of $G_2$, the 
combination of implicit effect of BC (\ref{Y2}) and explicit effect of 
BC (\ref{Y1}) can unambiguously determine the analytical behavior 
of $G_3$ along $y_1$. 
Similarly we can determine the analytical dependence of $G_1$ on 
$y_2$ without needs to impose the corresponding explicit boundary condition. 
The {\em implicit} power of
BCs (\ref{Y1}), (\ref{Y2}) obviously is effective only for differential 
equations on multiple variables, and this gives rise to  the 
``chained-entanglement'' feature observed here in the higher-dimensional space.

This striking feature of boundary conditions on the gauge symmetry 
can be employed further to determine the extent of gauge symmetry mixing 
in any number of space-time dimensions. But before doing that, let us quickly 
discuss a 7-dim 
set-up (i.e. 3-extra dim) to clearly show that gauge mixing/entanglement 
may indeed be 
destroyed by over-redundant BCs. Assume that we impose the following 
D-N BCs on 3-gauge symmetries in 7-dim
\bea
\nonu
&&\mbox{ along $Oy_1$:}
\\
&&
\label{3Y1}
\left\{
\begin{array}{c}
\partial_{y_1} G_2 |_{y_1=0} = \partial_{y_1} G'_2 |_{y_1=\pi R} =0 \\
G_3 |_{y_1=0} =  G'_3 |_{y_1=\pi R} =0 
\end{array}
\right. \;\;\;
\mbox{where} \;\;\;
\left(
\begin{array}{c}
G'_1 \\ G'_2 \\ G'_3
\end{array}
\right) \equiv
\left(
\begin{array}{ccc}
1&0&0\\
0&\cos \phi_1 & -\sin \phi_1  \\
0&\sin \phi_1 &\cos \phi_1 
\end{array}\right)
\left(
\begin{array}{c}
G_1 \\ G_2 \\ G_3
\end{array}
\right)
\eea
\bea
\nonu
&&\mbox{ along $Oy_2$:}
\\
\label{3Y2}
&&
\left\{
\begin{array}{c}
G_1 |_{y_2=0} =  G''_1 |_{y_2=\pi R} =0 \\
\partial_{y_2} G_3 |_{y_2=0} = \partial_{y_2} G''_3 |_{y_2=\pi R} =0 
\end{array}
\right. \;\;\;
\mbox{where} \;\;\;
\left(
\begin{array}{c}
G''_1 \\ G''_2 \\ G''_3
\end{array}
\right) \equiv
\left(
\begin{array}{ccc}
\cos \phi_2 & 0&\sin \phi_2  \\
0&1&0\\
-\sin \phi_2 &0&\cos \phi_2  
\end{array}\right)
\left(
\begin{array}{c}
G_1 \\ G_2 \\ G_3
\end{array}
\right)
\eea
\bea
\nonu
&&\mbox{ along $Oy_3$:}
\\
\label{3Y3}
&&
\left\{
\begin{array}{c}
\partial_{y_3} G_1 |_{y_3=0} = \partial_{y_3} G'''_1 |_{y_3=\pi R} =0 \\
G_2 |_{y_3=0} =  G'''_2 |_{y_3=\pi R} =0 
\end{array}
\right. \;\;\;
\mbox{where} \;\;\;
\left(
\begin{array}{c}
G'''_1 \\ G'''_2 \\ G'''_3
\end{array}
\right) \equiv
\left(
\begin{array}{ccc}
\cos \phi_3 &- \sin \phi_3 &0  \\
\sin \phi_3 &\cos \phi_3 &0 \\
0&0&1  
\end{array}\right)
\left(
\begin{array}{c}
G_1 \\ G_2 \\ G_3
\end{array}
\right)
\eea
The separated solutions corresponding to each of the above BC set are
\bdm
\left\{
\begin{array}{l}
G_2\sim  \cos(\frac{m_1}{R}+\frac{\phi_1}{\pi R})y_1   \\
G_3\sim -\sin(\frac{m_1}{R}+\frac{\phi_1}{\pi R})y_1 
\end{array}
\right.
\;\;
\left\{
\begin{array}{l}
G_1\sim  -\sin(\frac{m_2}{R}+\frac{\phi_2}{\pi R})y_2 \\
G_3\sim  \cos(\frac{m_2}{R}+\frac{\phi_2}{\pi R})y_2
\end{array}
\right.
\;\;
\left\{
\begin{array}{l}
G_1\sim  \cos(\frac{m_3}{R}+\frac{\phi_3}{\pi R})y_3 \\
G_2\sim  -\sin(\frac{m_3}{R}+\frac{\phi_3}{\pi R})y_3
\end{array}
\right.
\edm
It is not difficult to convince ourselves that, indeed the above three 
separated solutions cannot be consolidated into a single one of the type 
(\ref{6sol}), i.e. no decomposition solution exists if we use altogether
three D-N BC sets (\ref{3Y1}),(\ref{3Y2}),(\ref{3Y3}). Again, this is 
because BC (\ref{3Y1}) entangles $G_2$ with $G_3$, (\ref{3Y2}) entangles 
$G_3$ with $G_1$, and at this point $G_1$ and $G_2$ should have been already 
mixed by virtue of $G_3$. Since we here pressed on to impose (\ref{3Y3}) which 
once more directly mixes $G_1$ and $G_2$, this new BC 
apparently is redundant and in conflict with the effect of 
(\ref{3Y1}),(\ref{3Y2}). 

The Dirichlet-Neumann BCs of the types 
(\ref{Y1}), (\ref{Y2}) have been argued in literature to be consistent 
with action's variational principle (e.g. \cite{CGMPT} for 5-dim, 
\cite{GNS} for higher dimensions) by showing that the non-redundant D-N
BC set is sufficient to nullify all surface terms in the variation of the 
action $\delta S$. Consequently, non-redundant D-N BC set can be properly 
obtained 
by minimizing the action (and there may be more than one 
qualified/alternative non-redundant BC sets  depending on the action 
parameters' values \cite{HMN,CGMPT}). As thus, 
by applying the
variational principle on action, we generally would not expect 
to be able to generate {\em all} BCs 
(e.g. (\ref{3Y1}),(\ref{3Y2}),(\ref{3Y3})) in redundant set  
because those BCs are more than enough/needed (i.e. redundant) to nullify 
the surface terms. 
The reason here again is the implicit power of each BC on all 
coordinates in higher dimensions, 
that creates conflicting effects on gauge fields 
wave functions if all BCs of redundant set are simultaneously imposed.
\subsection{General perspectives: $S$ symmetries mixed in $d$ extra dimensions}
\begin{figure}
        \begin{center}    
        \epsfig{figure=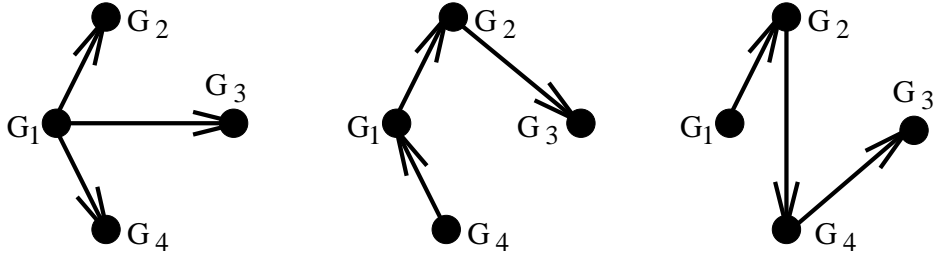,width=0.75\textwidth}
        \end{center}
        \vspace{0cm}
        \caption{Three of possible ``non-redundant'' ways to ``entangle''
the four gauge symmetries $G_1,G_2,G_3,G_4$. Each link (or arrow) represents
a set of 1D-1N boundary condition that mixes the two corresponding connected 
symmetries. Non-redundancy means that there is only one path connecting 
any two symmetries. The number of non-redundant links equals $(S-1)$ in 
graph with $S$ dots.}
        \label{noredundancy}
        \end{figure}

\begin{figure}
        \begin{center}    
        \epsfig{figure=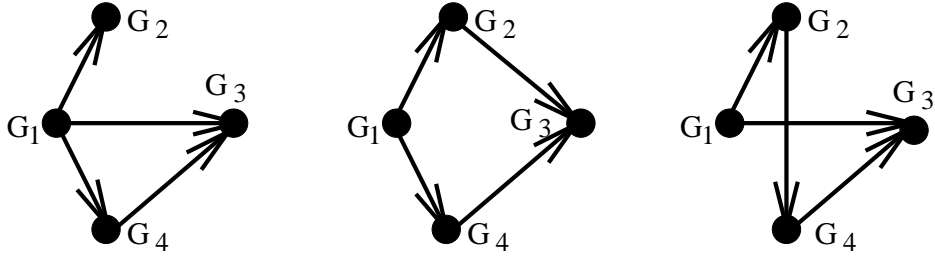,width=0.75\textwidth}
        \end{center}
        \vspace{0cm}
        \caption{Three of possible ``redundant'' ways to ``entangle''
the four gauge symmetries $G_1,G_2,G_3,G_4$. Redundancy means that there 
are more than one distinct paths connecting two symmetries. For e.g., in 
the first graph, two distinct paths connecting $G_1$ and $G_3$ are: 
($G_1 - G_3$) and ($G_1 - G_4 - G_3$). The redundacy will destroy the 
decomposition solution of the gauge fields.}
        \label{redundancy}
        \end{figure}

We are now ready for the general case of $S$ symmetry degrees of freedom in
$(d+4)$ space-time dimensions. It may be very useful if we visualize each 
symmetry as a dot, and each set of 1D-1N BC 
(e.g. (\ref{3Y1}) or (\ref{3Y2}) or (\ref{3Y3})) that entangles two symmetries 
as a link connecting the two corresponding dots.
In this picture, to avoid the above redundancy of boundary conditions, there 
should be only one possible (whether direct or indirect) path to travel 
between any two dots (along the established links) 
(Figure \ref{noredundancy}). 
If there are two distinct 
paths connecting dot $G_i$ to dot $G_j$, this just means gauge 
fields $G_i$ and $G_j$ 
are entangled by two independent ways using the given sets of BCs 
(Figure \ref{redundancy}). From the  
discussion of the last section, this ``multiple entanglement'' would 
destroy the mixed decomposition of gauge fields by the redundancy of those 
boundary conditions. 
According to this ``non-redundancy'' rule, 
in a graph with $S$ dots, we can establish $(S-1)$ ``non-redundant'' 
links, each represents a 1D-1N BC set in our convention. With these $(S-1)$ 
links, we also note that no dot (or a set of dots) is disconnected 
from the rest in the graph, which keeps the non-factorizability of the 
decomposition solution of the entire gauge fields' set.

Next we embed this ``chained entanglement'' into the $(d+4)$-dim space-time 
(i.e. $d$ extra dimensions). Each 1D-1N BC set along 1 extra dimension,
say $Oy_1$, will contribute two possible factors 
($\cos(\frac{m_1}{R}+\frac{\phi_1}{\pi R})y_1$ and 
$\sin(\frac{m_1}{R}+\frac{\phi_1}{\pi R})y_1$) to the gauge fields' wave 
functions. So in case of $d$ extra dimensions, the composite KK tower can 
accommodate/mix up to $2^d$ initial gauge symmetries
$\{G_1,\ldots,G_{2^d}\} $. More specifically,
the corresponding decomposition solution looks like
\bea
\nonu
&&G_{1}(x_{\mu},y_1,\ldots,y_d)=
\sum_{m_1,\ldots,m_d \in Z} g^{(m_1,\ldots,m_d)}(x_{\mu}) 
\cos(\frac{m_1}{R}+\frac{\phi_1}{\pi R})y_1 \ldots
\cos(\frac{m_d}{R}+\frac{\phi_d}{\pi R})y_d
\\
\label{many}
&&\ldots
\\
\nonu
&&G_{2^d}(x_{\mu},y_1,\ldots,y_d)=
\sum_{m_1,\ldots,m_d\in Z} g^{(m_1,\ldots,m_d)}(x_{\mu}) 
\sin(\frac{m_1}{R}+\frac{\phi_1}{\pi R})y_1 \ldots
\sin(\frac{m_d}{R}+\frac{\phi_d}{\pi R})y_d
\eea 

Combining the observations obtained in this section, we finally come to the 
following conclusion:

{\em In the space with $d$ extra finite dimensions, it is possible to mix 
up to 
$S\leq 2^d$ gauge symmetries by imposing $(S-1)$ sets of 1D-1N boundary 
conditions.}

There is one more subtle but interesting point underlying this conclusion 
that we need to address. That is, can all sets of BCs employed be totally 
independent? In general, the answer is negative, simply because if they were 
all different, then the ``entangled''  KK spectrum would be characterized by 
{\em up to} $(2^d-1)$ different twist angles $\{\phi_i\}$. But as 
we have seen in Eq. (\ref{many}) that the most extensive ``entangled'' spectrum
can contain only $d$  twist angles, so clearly BC sets cannot be all 
different in general. Below is a more rigorous analysis of this issue.
 
In the case of two extra dimension, we now have some basis to expect 
that we may mix four gauge symmetries $\{G_1,G_2,G_3,G_4\}$
(but not only three as we did in Eq. (\ref{6sol})). Here we attempt 
to do this with three different sets of 1N-1D BCs 
(i.e. $\phi_{11} \neq \phi_{12} \neq \phi_{2}$ {\em a priori})
\be
\label{3BCs}
G_{1,N} [O^{(y_1,\phi_{11})}(2)] G_{2,D}; \;\;\;\;\;\;\;
G_{3,N} [O^{(y_1,\phi_{12})}(2)] G_{4,D}; \;\;\;\;\;\;\;
G_{1,N} [O^{(y_2,\phi_{2})}(2)] G_{3,D}; 
\ee
where the short-hand notation 
$G_{1,N} [O^{(y_1,\phi_{11})}(2)] G_{2,D}$ denotes the familiar 1D-1N BC set 
imposed on $G_1,G_2$ along $y_1$-direction with twist angle $\phi_{11}$. 
The 
decomposition solution of each of these BC sets separately is
\bdm
\left\{
\begin{array}{l}
G_1\sim  \cos(\frac{m}{R}+\frac{\phi_{11}}{\pi R})y_1   \\
G_2\sim -\sin(\frac{m}{R}+\frac{\phi_{11}}{\pi R})y_1 
\end{array}
\right.
\;\;
\left\{
\begin{array}{l}
G_3\sim  \cos(\frac{m}{R}+\frac{\phi_{12}}{\pi R})y_1 \\
G_4\sim  -\sin(\frac{m}{R}+\frac{\phi_{12}}{\pi R})y_1
\end{array}
\right.
\;\;
\left\{
\begin{array}{l}
G_1\sim  \cos(\frac{m}{R}+\frac{\phi_2}{\pi R})y_2 \\
G_3\sim  -\sin(\frac{m}{R}+\frac{\phi_2}{\pi R})y_2
\end{array}
\right.
\edm
Clearly, when $\phi_{11}\neq \phi_{12}$ it is impossible to consolidate 
the above three solutions into a unified one. In fact, we can only do this if 
$\phi_{11} =\phi_{12} \equiv \phi_1$, and the unified KK decomposition 
satisfying all three BC sets (\ref{3BCs}) reads
\bea
\nonu
G_1(x_{\mu},y_1,y_2)&=&\sum_{m_1,m_2\in Z}g^{(m_1,m_2)}(x_{\mu})\;
\cos{\left([\frac{m_1}{R}+\frac{\phi_1}{\pi R}]y_1\right)}\;
\cos{\left([\frac{m_2}{R}+\frac{\phi_2}{\pi R}]y_2\right)}
\\
\label{4sol}
G_2(x_{\mu},y_1,y_2)&=&-\sum_{m_1,m_2\in Z}g^{(m_1,m_2)}(x_{\mu})\;
\sin{\left([\frac{m_1}{R}+\frac{\phi_1}{\pi R}]y_1\right)}\;
\cos{\left([\frac{m_2}{R}+\frac{\phi_2}{\pi R}]y_2\right)}
\\
G_3(x_{\mu},y_1,y_2)&=&-\sum_{m_1,m_2\in Z}g^{(m_1,m_2)}(x_{\mu})\;
\cos{\left([\frac{m_1}{R}+\frac{\phi_1}{\pi R}]y_1\right)}\;
\sin{\left([\frac{m_2}{R}+\frac{\phi_2}{\pi R}]y_2\right)}
\nonu
\\
G_4(x_{\mu},y_1,y_2)&=&\sum_{m_1,m_2\in Z}g^{(m_1,m_2)}(x_{\mu})\;
\sin{\left([\frac{m_1}{R}+\frac{\phi_1}{\pi R}]y_1\right)}\;
\sin{\left([\frac{m_2}{R}+\frac{\phi_2}{\pi R}]y_2\right)}
\nonu
\eea
Indeed, all four initial symmetries $\{G_1,G_2,G_3,G_4\}$ was 
non-trivially mixed in 6-dim to produce a single (composite) KK gauge field 
tower $g^{(m_1,m_2)}(x_{\mu})$ in 4-dim. The analysis also points out an 
important feature: for the existence of a unified decomposition solution, 
only one BC twist angle can be associated with each extra dimension, though
a same set of 1D-1N BCs can be imposed repeatedly on more than one set of 
fields. (In above e.g., when $\phi_{11} =\phi_{12}$, the same BC set indeed
was imposed on both $(G_1-G_2)$ and $(G_3-G_4)$). 

Finally, we note that the construction of gauge symmetry mixing 
in higher dimension presented here is not the 
the only possible or most-efficient one, because there are more possible ways 
to define BCs (e.g. on surface instead of on a direction) with more 
dimensions. 
This construction however has quite simple and interesting entanglement 
structure because it is built on the limitation of symmetry 
mixing in one extra dimension found in section \ref{Factorizability}.
The construction further yields exact solutions (\ref{6sol}), (\ref{4sol}), 
which are sufficient for our current primary goal to show the 
explicit non-trivial ``chained entanglement'' 
of up to $2^d$ gauge symmetries in $(d+4)$-dim space-time. 
A general treatment of gauge fields decomposition in higher dimensions by 
any imposition patterns of Dirichlet-Neumann BCs however lies beyond the scope 
of this work. 
\section{Conclusion}
\label{Conclusion}
In this work we have investigated the symmetry breaking of arbitrary gauge 
group by general Dirichlet-Neumann boundary conditions imposed at two ends 
of a fifth dimensional interval. Such boundary conditions induce the 
factorization of the initial gauge symmetry space into subspaces of only 
one or two dimensions. Those subspaces are mutually orthogonal to one 
the other, and thus each gives an independent KK tower of gauge fields 
in 4-dim. 

In gauge space, each unbroken symmetry direction form such one-dimensional 
``decoupled'' subspace, which is necessarily 
represented by a vector field having Neumann boundary conditions at both 
interval's ends. Each two-dimensional ``entangled'' subspace is 
characterized by a twist angle $\phi$, which gives the relative orientation 
between field sets with defined BCs at end-points. This angle determines 
the mass spectrum of the associated 4-dim tower, and can be tuned to produce 
phenomenological light gauge boson as its lowest mode. Further, in the 
4-dim effective picture, 
while this angle has no effect on the modes' diagonality and normalization, 
it controls the non-Abelian triple and quartic gauge boson self-interaction.

In an intact 5-dim viewpoint, though the group symmetry breaking scheme 
makes sense locally at end-points, 
it does not at any other points in the bulk. Matter fields with definite 
charges under end-groups then can be placed at end-branes to calibrate 
brane/bulk coupling ratio, or else the fifth coordinate can be 
integrated out to produce a full 4-dim effective picture. 

In higher dimension, the imposition of Dirichlet-Neumann BCs on pairs 
of gauge fields along 
different finite dimensions creates an very interesting 
``chained entanglement'' of the symmetries. This entanglement is also very 
strict in its nature that it tolerates just a minimum (non-redundant) sets of 
1D-1N BCs. 
We hope to come back with a more general analysis of gauge symmetry breaking 
on higher and non-flat intervals in future.   
\begin{acknowledgments}
The work is dedicated to the memory of Prof. Henri Van Regemorter, formerly 
at the Observatoire Paris-Meudon, France. 
\end{acknowledgments}
\begin{appendix}
\section{A lemma on matrix factorization}
\label{Alemma}
In this appendix, we will prove a lemma, which constitutes the formal 
ground for the factorizability of gauge symmetry space by 
Dirichlet-Neumann presented in section \ref{Factorizability}. 

Lemma: {\em Any general orthogonal matrix $O(2N)$ can always be 
brought into the block-diagonal form (of $N$ $O(2)$-blocks) by four  
separate and independent $O(N)$ rotations, two of which act on the right 
and the other two act on the left of the original $O(2N)$ matrix.}

In expression, this means, given a general $2N \times 2N$ orthogonal matrix 
$O \equiv $ $\left(
\begin{tabular}{l|r}
M&N \\ \hline
P&Q
\end{tabular} \right)$, 
one can always find four independent $O(N)$ matrices $A,B,C,D$ for which 
the following equality holds
\be 
\label{lemma}
\left(
\begin{tabular}{l|r}
A&0 \\ \hline
0&B
\end{tabular} \right) 
\left(
\begin{tabular}{l|r}
M&N \\ \hline
P&Q
\end{tabular} \right) 
\left(
\begin{tabular}{l|r}
C&0 \\ \hline
0&D
\end{tabular} \right) =
\left(
\begin{tabular}{cccc|cccc}
$c_{1} $&0&&0&$s_{ 1}$&0&&0 \\
0&$c_{2}$&&0&0&$s_{2}$&&0 \\
&&$\ddots$&&&&$\ddots$& \\
0&0&&$c_{N}$&0&0&&$s_{N}$ \\ \hline
$-s_{1}$ &0&&0& $c_{1}$ &0&&0 \\
0&$-s_{2}$&&0&0&$c_{2}$&&0 \\
&&$\ddots$&&&&$\ddots$& \\
0&0&&$-s_{N}$&0&0&&$c_{N}$ 
\end{tabular} \right) 
\ee 
where $c_{i} \equiv \cos{\phi_i}$, $s_{i} \equiv \sin{\phi_i}$. Evidently, 
the matrix on the right hand side has the block-diagonal form 
(of $N$ $O(2)$-blocks) after some permutations of columns (and rows).

Proof: Given the matrix $O \in O(2N)$, our objective is to identify  
four matrices $A,B,C,D \in O(N)$ that satisfy Eq. (\ref{lemma}).  
Because $O$ is an orthogonal matrix, 
$OO^T=O^T O= {\bf 1}_{2N\times 2N}$ which implies relations between $M,N,P,Q$
\be
\label{MNPQ}
\left(\begin{tabular}{c|c}
$MM^T + NN^T$ & $MP^T + NQ^T$ \\ \hline
$PM^T + QN^T$ & $PP^T + QQ^T$
\end{tabular} \right) =
\left(\begin{tabular}{c|c}
${\bf 1}_{N\times N}$ & 0 \\ \hline
0 & ${\bf 1}_{N\times N}$
\end{tabular} \right) =
\left(\begin{tabular}{c|c}
$M^T M + P^T P$ &$ M^T N + P^T Q $\\ \hline
$N^T M + Q^T P$ &$ N^T N + Q^T Q $
\end{tabular} \right)
\ee
Let us define $A$ and $C^T$ as the $N\times N$ orthogonal matrices 
that diagonalize the symmetric matrices $MM^T$ and $M^T M$ respectively, i.e.
\be
\label{defAC}
A(MM^T)A^T= 
\left(\begin{array}{ccc}
c_1^2 & & \\
&\ddots& \\
&&c_N^2
\end{array} \right)
=C^T (M^T M)C
\ee
Because $M M^T + N N^T  = {\bf 1}_{N\times N} = M^T M + P^T P$ 
(see Eq. (\ref{MNPQ})), the above definitions of $A$ and $C$ also imply
\be
\label{NP}
A(NN^T)A^T = 
\left(\begin{array}{ccc}
s_1^2 & & \\
&\ddots& \\
&&s_N^2
\end{array} \right)
=C^T (P^T P)C
\ee
Next, let us define $B$ and $D^T$ as the orthogonal matrices that 
diagonalize $PP^T$ and $N^T N$ respectively (note that $KK^T$ and $K^T K$ have 
same set of eigenvalues for any square real matrix $K$)
\be
\label{defBD}
B(PP^T)B^T= 
\left(\begin{array}{ccc}
s_1^2 & & \\
&\ddots& \\
&&s_N^2
\end{array} \right)
=D^T (N^T N)D
\ee
And since $Q Q^T + P P^T  = {\bf 1}_{N\times N} = Q^T Q + N^T N$, 
in place of (\ref{NP}) now we have
\be
\label{BQD}
B(QQ^T)B^T = 
\left(\begin{array}{ccc}
c_1^2 & & \\
&\ddots& \\
&&c_N^2
\end{array} \right)
=D^T (Q^T Q)D
\ee
We first consider the non-degenerate case where $c_i \neq c_j$ for every 
$i\neq j$. Then the matrices $A,C$ in (\ref{defAC}) and $B,D$ in 
(\ref{defBD}) are uniquely defined. In consequence, from (\ref{defAC}) 
follows that the general real $N\times N$ matrix $M$ is diagonalized by 
$A$ and $C$, i.e. $(AMC)\sim diag(c_1,c_2,\ldots,c_N)$. Similar conclusions 
can be drawn from combination of (\ref{NP}) and (\ref{defBD}), or 
(\ref{defAC}) and (\ref{BQD}), so altogether we have
\be
\label{c}
|AMC|=   \left(\begin{array}{ccc}
|c_1| & & \\
&\ddots& \\
&&|c_N|
\end{array} \right) =|BQD|
\ee 
\be
\label{s}
|AND|=   \left(\begin{array}{ccc}
|s_1| & & \\
&\ddots& \\
&&|s_N|
\end{array} \right) =|BPC|
\ee 
\end{appendix}
Next, it follows from the relation $MP^T + NQ^T =0$ (see (\ref{MNPQ})) that
\be
\label{four}
\begin{array}{c}
0=A(MP^T + NQ^T)B^T=(AMC)(C^T P^T B^T)+ (AND)(D^T Q^T B^T) \Rightarrow \\
(AMC)(BPC)^T+(AND)(BQD)^T=(AMC)(BPC)+(AND)(BQD)= 0
\end{array}
\ee 
because both $(BPC)$ and $(BQD)$ are diagonal (\ref{c},\ref{s}). After 
explicitly expanding the matrix product on the left hand size of 
(\ref{lemma})
\be
\left(
\begin{tabular}{l|r}
A&0 \\ \hline
0&B
\end{tabular} \right) 
\left(
\begin{tabular}{l|r}
M&N \\ \hline
P&Q
\end{tabular} \right) 
\left(
\begin{tabular}{l|r}
C&0 \\ \hline
0&D
\end{tabular} \right) = 
\left(
\begin{tabular}{l|r}
AMC&AND \\ \hline
BPC&BQD
\end{tabular} \right) 
\ee 
and using (\ref{c}), (\ref{s}), (\ref{four}) we indeed obtain the equality 
(\ref{lemma}).

For the degenerate case (where $c_i = c_j$ for some $i \neq j$), all 
orthogonal matrices $A,B,C,D$ can only be determined up to some 
sub-rotation within each degenerate subspace. Due to this ambiguity, 
(\ref{defAC}) does not necessarily imply that $(AMC)$ be a diagonal matrix 
(and same for $(BQD)$, $(AND)$, $(BPC)$). However, as the degeneracy pattern 
is identical for (\ref{defAC}), (\ref{NP}), (\ref{defBD}), (\ref{BQD}) we 
can still pick simultaneously a set of $A,B,C,D$, 
for which all four matrices $(AMC)$, $(BQD)$, $(AND)$, $(BPC)$ are diagonal.  
In the result, in this case too, the equality (\ref{lemma}) is satisfied. 

In the Dirichlet-Neumann physical system considered in 
section \ref{Factorizability}, the 
$2N \times 2N$ orthogonal matrix $O$ presents the relation between two sets 
of $2N$ gauge fields with defined BCs at $y=0$ and $y=\pi R$ respectively. 
The $N \times N$ orthogonal matrices $A,B,C,D$ implement the allowed basis 
transformations within sets of $N$ gauge fields of the same (Dirichlet 
or Neumann) BC type. The equation (\ref{lemma}) then asserts that 
such $2N$-dim 
system can always be factorized into $N$ two-dimensional sub-systems (each is 
characterized by an $O(2)$-rotation angle $\phi_i$), whose KK decomposition 
is presented in section \ref{Propagation}. In addition, we specially note 
that (as it is evident from (\ref{defAC}), (\ref{NP}), (\ref{defBD}), 
(\ref{BQD})) the basis rotations characterized by $A,B,C,D$ do not have 
any effect on the set $\{\phi_1,\ldots,\phi_N\}$. Those angles are the only 
relevant parameters encoded in (and unique to) the initial set of BCs. 

\end{document}